\documentclass[acmlarge,natbib=false,manuscript,screen,nonacm]{acmart}
\setcopyright{none}
\acmDOI{}
\acmISBN{}
\acmConference[]{}{}{}
\renewcommand{\footnotetextcopyrightpermission}[1]{}

\usepackage{graphicx}
\usepackage{cleveref}
\crefname{figure}{Figure}{Figures}
\Crefname{figure}{Figure}{Figures}

\crefname{table}{Table}{Tables}
\Crefname{table}{Table}{Tables}

\usepackage[hypcap=true]{caption}
\usepackage{subcaption}
\usepackage{geometry}
\usepackage{array}
\usepackage{booktabs} % For better horizontal rules
\usepackage{array}
\usepackage{url}
\usepackage{multirow}
\usepackage{booktabs}
\usepackage{tabularx}
\usepackage{lscape}     % for landscape orientation
\usepackage{longtable}  % for tables spanning multiple pages

\AtBeginDocument{%
  \providecommand\BibTeX{{%
    \normalfont B\kern-0.5em{\scshape i\kern-0.25em b}\kern-0.8em\TeX}}}

\RequirePackage[
  datamodel=acmdatamodel,
  style=acmnumeric,
  ]{biblatex}

\begin{document}

\title[Android App Feature Extraction]{Android App Feature Extraction: A review of approaches for malware and app similarity detection}

\author{Simon Torka}
\authornotemark[1]
\email{simon.torka@dai-labor.de}
\orcid{0009-0009-1201-1915}
\affiliation{
  \institution{DAI-Lab, Technische Universität Berlin}
  \streetaddress{Ernst-Reuter-Platz 7}
  \city{Berlin}
  \state{Berlin}
  \country{Germany}
  \postcode{10587}
}
\author{Sahin Albayrak}
\authornotemark[2]
\email{sahin.albayrak@dai-labor.de}
\orcid{0000-0001-5092-4584}
\affiliation{
  \institution{DAI-Lab, Technische Universität Berlin}
  \streetaddress{Ernst-Reuter-Platz 7}
  \city{Berlin}
  \state{Berlin}
  \country{Germany}
  \postcode{10587}
}

% \renewcommand{\shortauthors}{Torka et al.}

\begin{abstract}
    This paper reviews work published between 2002 and 2022 in the fields of Android malware, clone, and similarity detection. It examines the data sources, tools, and features used in existing research and identifies the need for a comprehensive, cross-domain dataset to facilitate interdisciplinary collaboration and the exploitation of synergies between different research areas. Furthermore, it shows that many research papers do not publish the dataset or a description of how it was created, making it difficult to reproduce or compare the results. The paper highlights the necessity for a dataset that is accessible, well-documented, and suitable for a range of applications. Guidelines are provided for this purpose, along with a schematic method for creating the dataset.
\end{abstract}


%\received{31 Jan 2023}
%\received[revised]{29 Apr 2024}
%\received[accepted]{X}

\maketitle

\section{INTRODUCTION}
\label{sec:intro}

Smartphones are becoming increasingly important and indispensable for people’s daily lives because they offer a range of convenient features and functions that simplify everyday tasks. However, this fact leads to a dependency between users and their smartphones which can be dangerous if compromised applications are used. 

According to Ericsson (cited by \cite{Turner.2023}), in 2021, 6.34 billion smartphones were in use. The most widespread smartphone operating system is Google's Android, with at least 71.45\% market coverage, followed by Apple's iOS, with about 27.83\% market coverage \cite{Statista.28.07.2022}. The popularity of Android also has its drawbacks. According to \citeauthor{Willems.2022}~\cite{Willems.2022} from GDATA, currently, more than 25 million malicious applications for Android have been detected. Furthermore, not only intentionally manipulated apps can harm users. Poorly programmed or outdated applications can also pose significant risks. According to atlasVPN \cite{Ruth.28.07.2022}, "63\% of Android applications had known security vulnerabilities in Q1 2021, with an average of 39 vulnerabilities per app", whereby gaming and finance apps have the most security issues.

Malware is often distributed by adding malicious code to pirated apps, also known as "piggybacking".\footnote{According to \citeauthor{Zhou.2012}~\cite{Zhou.2012}, pirated apps are not only used to inject malicious code; They can also be used to “replace existing in-app advertisements or embed new ones to ‘steal’ or redirect advertising revenue” \cite[317]{Zhou.2012}. In addition to the threat posed by pirated apps due to inserted malicious code, they also represent a violation of applicable copyright law.} One way to identify piggybacking is to detect applications that are nearly identical but differ in minor ways.\footnote{This method can also detect currently unknown malicious code.} This approach demonstrates the advantages of combining similarity analysis and malware detection. Such strategies are essential to contemporary threat detection systems and require further investigation. Additionally, research into comprehensive systems that also consider the human factor as a potential security risk\footnote{Users may ignore risks and continue to use malicious applications if suitable alternatives are not provided.} is vital to the success of contemporary protection systems.

A fundamental requirement for conducting research on such systems is the availability of appropriate training datasets. However, the existing research landscape reveals deficiencies in the provision of this data.\footnote{The deficiencies include unpublished or no longer available datasets, errors in the description of the dataset generation process, and the omission of all dataset-relevant information. In addition, publicly available datasets usually focus on the area of malware detection and are often outdated, e.g., Android Genom Project \cite{Zhou.2012b}, M0DROID \cite{Damshenas.2015}, Drebin \cite{Arp.February23262014, Spreitzenbarth.2013}, Kharon \cite{Kiss.2016}, AAGM \cite{Lashkari.2017}, Android PRAGuard \cite{Maiorca.2015}, AndroZoo \cite{Allix.2016c} or unavailable, e.g., ContagioDump, AndroMalShare, AMD Project.} This makes it difficult to compare, evaluate, and reproduce research results. Furthermore, research on systems for detecting functionally similar applications and research on holistic systems is currently not possible because of the lack of universal datasets. 

To address this, we examine the literature on malware, app similarity, and clone detection published between 2002 and 2022 to identify all features suitable for creating such a dataset as well as the tools necessary to extract them. A keyword and snowballing search was performed to identify relevant literature\footnote{The keyword search was performed on "ACM digital library", "Google Scholar", and "IEEE Xplore" and included terms such as "Android + Similarity", "Clone + Android" and "Malware + Android". For the snowballing search, connectedpapers was used (https://www.connectedpapers.com/). This is a website to generate citation graphs based on the provided starting literature.}.

Furthermore, existing survey papers in the domains that have analyzed literature related to malware, app similarity, and clone detection were examined. However, these surveys often lack a comparative analysis between these domains, as well as a thorough discussion on the creation of derived datasets from identified features. For instance, \citeauthor{Li.2017}~\cite{Li.2017} focus on clone and malware detection but do not delve into similarity detection. \citeauthor{Firdaus.2018}~\cite{Firdaus.2018} explore static features extracted from app code and find the best feature set for malware detection by using a genetic algorithm but do not provide a comparative analysis of their application in clone and similarity detection. \citeauthor{Alzubaidi.2021}~\cite{Alzubaidi.2021} and \citeauthor{Feizollah.2015}~\cite{Feizollah.2015} also primarily concentrate on feature extraction for malware detection, neglecting similarity analysis. \citeauthor{Dhalaria.2021}~\cite{Dhalaria.2021} and \citeauthor{Jusoh.2021}~\cite{Jusoh.2021} provide detailed insights into malware detection techniques but do not discuss their results in the context of similarity detection. \citeauthor{Sharma.2021}~\cite{Sharma.2021} review 380 papers, offering a comprehensive overview of features and techniques but solely within the context of malware detection. \citeauthor{Selvaganapathy.2021}~\cite{Selvaganapathy.2021} also focus on the area of malware detection. In their work, they provide a list of public datasets for malware detection and briefly discuss approaches to feature analysis and the tools most commonly used for this purpose.

Our research expands the existing literature by jointly examining the key research areas that have been analyzed individually so far. In a comprehensive literature mining process, features usable across disciplines are identified and analyzed in detail. This forms the basis for a comprehensive multi-purpose dataset. In addition, we aim to define the requirements that such a dataset must fulfill and outline the process for its creation.

This paper is structured as follows: section 2 lays the technical foundations of Android application analysis and examines the mechanisms underlying these applications. In addition, sources for collecting applications are shown and tools that can be used to capture features are discussed. In section 3, different Android features extractable from the app stores or the APK files are introduced. It is also discussed how these features are used in different domains. Section 4 describes the process of generating a universal, interdisciplinary dataset, and forms the core of this work together with the analysis of extractable features. Section 5 critically discusses the limitations of this work. Finally, section 6 offers a reflection of the findings and implications from the previous chapters.
\section{THE TECHNICAL FOUNDATION OF ANDROID APPLICATION ANALYSIS}
\label{sec:technical_foundation}

This section lays the foundation for this survey paper by providing information on the technical background of Android and Android application analysis tools. All essential information starting from the definition of how an Android application is built up, executed, and distributed, as well as information about sources to gather Android applications, tools to extract data from Android applications, and features that can be extracted from Android applications, are presented.

\subsection{EXPLORING THE MECHANISMS UNDERLYING ANDROID APPLICATIONS}
\label{ssec:function_android}

\begin{figure}[ht]
    \centering
    \includegraphics[width=0.5\textwidth]{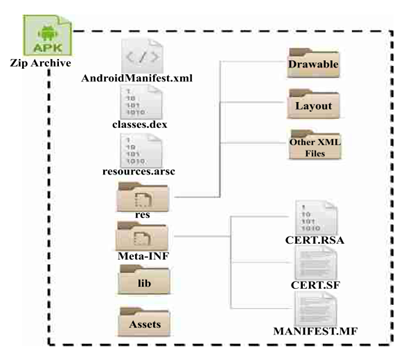}
    \caption{APK inner structure by \citeauthor{Hamednai.2019}~\cite{Hamednai.2019}}
    \label{fig:Fig_1_APK_Stucture_Hamednai_2019}
    \Description{A schematic of the internal structure of the .apk zip archive, including AndroidManifest.xml, classes.dex, resources.arse, the res folder and the include drawable, layout and .xml folders, the meta-in folder, and the internal cert.rsa, cert.sf, and manifest.mf files, and the lib and assets folders.}
\end{figure}

Android’s operating system is a mobile device multi-user computing platform \cite{Chen.2013}. It was introduced by Andy Rubin and taken over by Google in 2005 \cite{Google.2022}. It relies on an open-source Linux kernel optimized for mobile usage and provides a separate virtual environment for each running application to increase security \cite{Schmidt.2009d}. Furthermore, an application framework and some core applications are supplied by Android. To expand and individualize the usability of Android, users can expand the function set of their smartphones by installing applications (Apps) for different purposes. These applications - mostly provided by third parties - can be downloaded from different stores and come in the format of an Android application file (APK). An .apk-file is similar to a Java .jar file. Jar files, and .apk-files, are compressed zip archives that contain the application resources and code \cite{Sahs.2012, Shabtai.2010}. After decompressing, the extracted folder contains the AndroidManifest.xml file, the classes.dex file, the resources.arsc file as well as folders for resources (res-folder), meta information (Meta-INF-folder), libraries (lib-folder), and assets (Asset-folder) (see  
\cref{fig:Fig_1_APK_Stucture_Hamednai_2019}), which can be used to derive a lot of useful information \cite{Gonzalez.2015, Hamednai.2019, Potharaju.2012, Xu.2018}.

The \textbf{\textit{AndroidManifest.xml}} holds meta-information about the application, for example the application’s name, version, the codes package (or namespace), and the permissions necessary for execution \cite{Crussell.2012, Hamednai.2019, Sahs.2012}. It also contains information about the app’s behavior, like its entry points \cite{Gonzalez.2016, Gonzalez.2015}. The \textbf{\textit{classes.dex}} file includes the application’s Dalvik byte code and can also be used to mine some helpful information. A developer must write the application’s code in Java to generate this dex-code. This Java application must be translated into Java byte code (.jar). Because Android smartphones cannot interpret Java byte code, a further translation step is necessary by translating the .jar-files into the Dalvik executable (DEX) format \cite{Chen.2013, Crussell.2012, Gonzalez.2015, Hanna.2013}. The DEX format is no longer human-readable and can only be interpreted by the Dalvik virtual machine \cite{Crussell.2012, Zhou.2012} or with the Android runtime (ART) \cite{Hamednai.2019}. To investigate Android security, gathering information from the dex code and transforming it back into several intermediate languages or to the original Java format is possible. The precompiled application resources are held in the \textbf{\textit{resources.arsc}} \cite{Hosseini.2021}. The \textbf{\textit{res-folder}} contains all data which are not compiled into the resources.arsc folder, e.g., sounds, images, icons, and xml files necessary to run the application, as well as its UI layouts \cite{Crussell.2012, Gonzalez.2016, Gonzalez.2015, Hamednai.2019}, and could also be used as a data source. The \textbf{\textit{Meta-Inf-folder}} contains the digital signature of the application as well as the author.\footnote{In detail the authentication is done with the CERT.RSA file, CERT.SF files as well as the MANIFEST.MF file.} This information is necessary to identify the author and validate the authenticity of the ownership \cite{Zhou.2012}. It contains the developers’ names, their contact and organization information, as well as the public key fingerprints for validation \cite{Gonzalez.2015, Hamednai.2019, Zhou.2012}. It can be seen as the authors’ base information which can also be used to identify different security-related aspects. The \textbf{\textit{lib-folder}} contains all compiled executables and all native libraries as lib.so files. This code can be called from the application’s .dex-file using the Java native interface (JNI) \cite{Gonzalez.2016, Hosseini.2021}. It can be used to compare apps and detect malicious data injected into the application during runtime. The application’s assets like the FAQ or license information are stored in the \textbf{\textit{assets-folder}} \cite{Hosseini.2021}. This information can also be used as a data source to detect Android security-related purposes. 

Because the \textbf{\textit{dex-file}} contains the application’s source code and thus the application’s behavior, this data will be investigated in more detail. To run a dex file, an interpreter is necessary. The Dalvik Virtual Machine, as part of the applications’ frameworks, can be used \cite{Chen.2013, Potharaju.2012}. Most dex files are third-party applications and can be downloaded from different Android application stores. To understand how Android works it is necessary to understand the inner structure\footnote{A detailed explanation of how a dex file is organized can be found in \cite{Lynch.2018} which is the base for the following chapter.} of a .dex-file which can be seen in \cref{tab:Fig_2_APK_inner_structure_Hosseini_2021}\footnote{A graph of all sections used in an APK can be found under https://formats.kaitai.io/dex/dex.svg.}. According to \citeauthor{Lynch.2018}~\cite{Lynch.2018}, the .dex-file is organized into several sections. The first section is the \textbf{\textit{header-section}} which contains some meta-data about the application. A detailed view of the header file is given in \cref{tab:Tab_1_inner_struct_of_header_lynch_2018}.

\vspace*{15px}
\begin{minipage}[b]{0.4\linewidth}
    \centering
    \includegraphics[width=\linewidth]{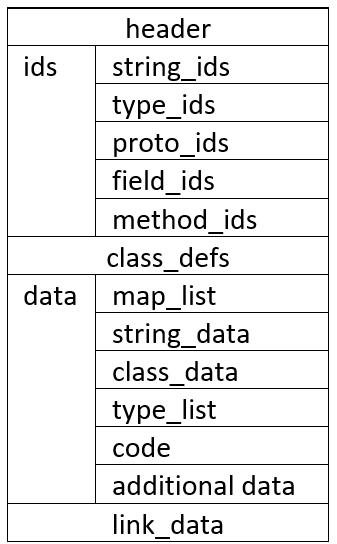}
    \captionof{figure}{The inner structure of a .dex-file. Own visualization based on \citeauthor{Hosseini.2021}~\cite{Hosseini.2021}}
    \label{tab:Fig_2_APK_inner_structure_Hosseini_2021}
    \label{fig:apk_structure}
    \Description{The inner .dex structure starting with the header, the ids (including string_ids, type_ids, proto_ids, field_ids and metod_ids), the class_defs, the data (including map_list, string_data, class_data, type_list, code and additional data) and the link_data.}
\end{minipage}
\hspace*{15px}
\begin{minipage}[b]{0.4\linewidth}
    \centering
    \captionof{table}{Inner structure of the header field. Own visualization based on \citeauthor{Lynch.2018}~\cite{Lynch.2018}}
    \label{tab:Tab_1_inner_struct_of_header_lynch_2018}
    \begin{tabular}{p{3cm}p{4cm}}
    \hline
    \multicolumn{1}{c}{\textbf{Field Name}} & \multicolumn{1}{c}{\textbf{Description}} \\
    \hline
    DEX\_FILE\_MAGIC & unique file type identifier \\
    Adler-32 Checksum & checksum for validation \\
    SHA-1 HASH & used as identifier e.g. in multidex scenarios \\
    File Size & size of the APK in bytes \\
    Header Size & size of the header information in bytes \\
    Endian Tag & Identifier for big or little endian coding \\
    Link Size & size of library link id field \\
    Link Offset & offset to find the field \\
    Map Offset & offset to find the field \\
    String Ids Size & size of the string id field \\
    String Ids Offset & offset to find the field \\
    Type Ids Size & size type of id field \\
    Type Ids Offset & offset to find the field \\
    Proto Ids Size & size of the proto id field \\
    Proto Ids Offset & offset to find the field \\
    Field Ids Size & size of the field id field \\
    Field Ids Offset & offset to find the field \\
    Method Ids Size & size of the method id field \\
    Method Ids Offset & offset to find the field \\
    Class Defs Size & size of the class def field \\
    Class Defs Offset & offset to find the field \\
    Data Size & size of application data \\
    Data Offset & offset to find application data \\
    \hline
\end{tabular}

\end{minipage}
\vspace*{15px}

As claimed by \citeauthor{Lynch.2018}~\cite{Lynch.2018}, the first entry of each header-section is the DEX\_FILE\_MAGIC, which is only an identifier to detect that the following file is a dex-file. The value is constantly called "dex", followed by the SDK’s target file version. The Checksum is for validating the given file. SHA-1 Hash can be used to identify .dex files in a multidex scenario. Next, the file size provides information about the size of the whole file, while the header-size gives only the size of the header. The endian tag is the last entry in the header and tells if big- or little-endian encoding is used.

Next, IDs for strings, types, protos, fields, methods, and classes are specified. According to \citeauthor{Lynch.2018}~\cite{Lynch.2018}, each ID field consists of the size of the referenced data and the reference address offset information, which keeps the required data. Furthermore, all data in these sections are sorted by their content and should not have any duplicate information \cite{Hamednai.2019}. 

By resolving the string IDs, textual information about classes, methods, parameters, variables, (error) messages, and user dialogues can be found \cite{Hamednai.2019}. The resolved type IDs identify all types used in the application, like classes, arrays, or primitive data types. Therefore, the types reference the string fields to specify the type names. The resolved proto IDs determine function prototypes by clarifying their return types and the number of needed parameters\footnote{A list of types and more information about generating prototypes can be found in \cite{AndroidOpenSourceProject.2022}.}. The fields section contains information about all defined data fields used in the application. The method section defines each user-defined method as well as all called API methods. Each method definition consists of three parts. The method’s name references the string table, prototype references, defined prototypes, and class ID. The last entry in the IDs section is the class section, where all used classes are defined. The mechanism behind those definitions is the same as the one behind the function referencing. 

All code and data representing the application are stored in the data section. The previously introduced ID-sections reference all their content in this data section by offset and size. Therefore, this section consists of the content of the other sections and can be easily accessed using the underlying addresses \cite{Gonzalez.2016, Hamednai.2019}.

The link data section is the last section of a .dex-file and contains information related to all static linked native libraries \cite{Lynch.2018}.

Sometimes it is necessary to analyze a dex-file (e.g., for security analysis or to detect app similarities). To do this, the non-readable dex-file must be transferred into a human-readable format. An option is to translate the .dex-file into the human-readable Smali format \cite{Felt.2011, Zhang.2018}. 

According to \citeauthor{McLaughlin.2017}~\cite{McLaughlin.2017}, when translating a dex-file into the human-readable Smali format, each class contained in the dex-file is transferred to a separate Smali-file. In addition, each Smali file contains the methods of the underlying class as well as their instructions and operands, in a human-readable format. According to \citeauthor{Hosseini.2021}~\cite{Hosseini.2021}, all instructions in a method can be interpreted as a sequence of instructions representing the flow of the function.

But why is all that necessary, why can Android apps be dangerous, and how can malware be injected into Android applications? The answer lies in the concept of how the Android ecosystem works. As mentioned, most Android applications are provided by third-party authorities. The process of publishing and distributing Android applications is a gateway for malware. According to \citeauthor{Peng.2012}~\cite{Peng.2012}, developers need a publisher account to distribute their Android applications in the Play Store. Only a Gmail account and an initial payment of 25.00 USD are required \cite{Potharaju.2012}. After creating an account, each developer gets a digital certificate. According to \citeauthor{Potharaju.2012}~\cite{Potharaju.2012}, it is necessary to sign an application before uploading it to the Play Store with the author’s public key. Then the application can be uploaded and get its product page at the Play Store \cite{Peng.2012}. According to \citeauthor{Potharaju.2012}~\cite{Potharaju.2012}, each product site delivers meta information like the app’s name, the users’ rating of the application, the update history, the current app’s version, the category to which the app belongs, the number of installs, the size of the application and the application’s price. Furthermore, there are some Android stores from other companies than Google like apkpure \cite{APKPure.10.02.2021}. Because of this concept, it is really easy to clone applications, insert malware, and republish them on the same or other stores \cite{Crussell.2012}.

\subsection{STRATEGIES FOR COLLECTING ANDROID APPLICATIONS}
\label{ssec:gather_andorid_apps}

Sources to gather Android applications are manifold. However, according to \citeauthor{Schmicker.2019}~\cite[67]{Schmicker.2019}, this process "can be time consuming, e.g., one may be tasked with writing a crawler or contacting website administrators for access to a bulk download". Furthermore, some authors do not describe the composition of their data well. The problems range from unspecified app sources to insufficiently documented or incomplete data. Furthermore, some datasets are no longer available or have not been published. The unavailability of some 3\textsuperscript{rd} party stores further complicates data access. For this paper, all sources are extracted from the analyzed literature and listed in \cref{tab:Tab_2_Android_APK_Sources}\footnote{Datasets that are not described well enough and unavailable 3\textsuperscript{rd} party stores are not considered in \cref{tab:Tab_2_Android_APK_Sources}. For example, the referenced dataset from \citeauthor{Zhou.2013}~\cite{Zhou.2013} is not available, and the one from \citeauthor{Schmicker.2019}~\cite{Schmicker.2019} is no longer publicly accessible. In the case of \citeauthor{Sahs.2012}~\cite{Sahs.2012}, the dataset cannot be mentioned because insufficient information about it is available (the app source is not specified). In \citeauthor{Shabtai.2010}~\cite{Shabtai.2010}, only the data source ([official] Android Market) and statistical information about apps are given, but not the names of these apps. In \citeauthor{LinaresVasquez.2016}~\cite{LinaresVasquez.2016}, the data source (Play Store) and a list of app names and categories are available but not the app version, the APK or a dataset with all extracted features. In these cases, the original data source (Play Store) is listed in \cref{tab:Tab_2_Android_APK_Sources} but not the derived dataset.}. 

The data are split into three domains: "Official Android Store", which is the Google Play Store, "3rd Party Android App Store", which are all other stores to gather Android applications, and "Datasets", which are available to analyze Android applications and malware. Crawlers can be used to collect applications from the Play Store or 3rd Party stores. Some authors use VirusTotal after crawling the applications to decide if they are benign or malicious.

\begin{table}[ht]
    \centering
    \caption{Data Sources for Android Applications}
     \label{tab:Tab_2_Android_APK_Sources}
    \begin{tabular}{p{1.3cm}p{5cm}l}
        \toprule
        \textbf{Type} & \textbf{Data source} & \textbf{URL} \\
        \midrule
        \textbf{Official} & Play Store & \url{https://play.google.com/} \\
        \midrule
        \textbf{3rd Party} & 1mobile & \url{https://1mobile-market.de.aptoide.com/app} \\
                                              & 9app & \url{https://www.9apps.com/} \\
                                              & Amazon Appstore for Android & \url{https://www.amazon.de/gp/mas/get/amazonapp} \\
                                              & ApkPure & \url{https://m.apkpure.com/} \\
                                              & AppChina & \url{http://m.appchina.com/} \\
                                              & Appland & \url{https://www.applandinc.com/} \\
                                              & appsapk & \url{https://www.appsapk.com/} \\
                                              & Aptoide & \url{https://de.aptoide.com/} \\
                                              & BAIDU & \url{http://shouji.baidu.com/} \\
                                              & Dangle Store & \url{http://android.d.cn} \\
                                              & Fdroid & \url{https://f-droid.org/} \\
                                              & freewarelovers & \url{https://www.freewarelovers.com/} \\
                                              & GetJar & \url{https://www.getjar.com/} \\
                                              & goapk & \url{https://goapk.org/} \\
                                              & Mumayi & \url{http://www.mumayi.cn/} \\
                                              & MYAPP & \url{http://android.myapp.com/} \\
                                              & proAndroid & \url{https://www.proandroid.com/} \\
                                              & Sliedme Store & \url{http://slideme.org/} \\
                                              & softportal & \url{https://www.softportal.com/} \\
        \midrule
        \textbf{Datasets} & UNB Android botnet dataset & \url{https://www.unb.ca/cic/datasets/android-botnet.html} \\
                           & UNB Android validation dataset & \url{https://www.unb.ca/cic/datasets/android-validation.html} \\
                           & GitHub Android Malware Collection & \url{https://github.com/ashishb/android-malware} \\
                           & Android Malware Genome Project & \url{http://www.malgenomeproject.org/} \\
                           & Dataset of Dienst and Berger & \url{http://www.informatik.uni-leipzig.de/~berger/tr/2012-dienst.pdf} \\
                           & Drebin & \url{https://drebin.mlsec.org/} \\
                           & DroidAnalytics & \url{http://ansrlab.cse.cuhk.edu.hk/software/droidanalytics/} \\
                           & Mobilewalla & \url{https://www.mobilewalla.com/} \\
                           & VirusShare Database & \url{https://virusshare.com/} \\
                           & Contagio Mobile & \url{https://contagiominidump.blogspot.com/} \\
                           & Das Malwerk & \url{http://dasmalwerk.eu/} \\
                           & theZoo & \url{https://thezoo.morirt.com/} \\                           
        \bottomrule
    \end{tabular}
\end{table}

\subsection{METHODOLOGIES FOR EXTRACTING FEATURES}
\label{sssec:extraction_process}

\begin{figure}[h]
    \centering
    \includegraphics[width=0.5\textwidth]{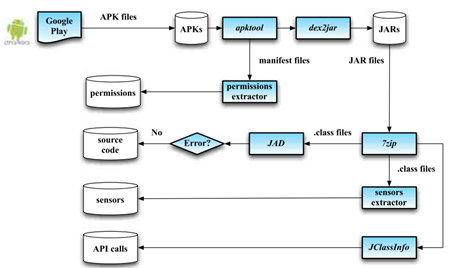}
    \caption{An example of a feature extraction toolchain \cite{LinaresVasquez.2016}.}
    \label{fig:Fig_3_Feature_Extraction_Toolchain_Linares_Vaquez_2016}
    \Description{Schematic toolchain starting with the .apk decompiled with APKTools, which extracts the permissions and provides the dex for the dex2jar tool. The dex2jar returns the jars which can be decompressed with 7zip into .class files. Then jad extracts the source code. The sensors can be extracted from the .class files. API calls can also be extracted using the JClassInfo tool.}
\end{figure}

For investigating Android applications, it is necessary to examine some valuable and analyzable features of the application. Many different tools were found, which are listed in \cref{tab:Tab_3_Feature_extraction_tools} and explained in this chapter.\footnote{An extended version of this table, containing information on all tools, including their input and output values, as well as their usability, can be found in the Appendix, \cref{tab:Tab_3_Feature_extraction_tools_extended}.} Most tools cannot be used stand-alone because they are specialized for specific purposes like decompiling, extracting, or analyzing. For this reason, toolchains are necessary. An illustration of a toolchain for analyzing Android applications can be found in \cref{fig:Fig_3_Feature_Extraction_Toolchain_Linares_Vaquez_2016} and was introduced by \citeauthor{LinaresVasquez.2016}~\cite{LinaresVasquez.2016}.
%Therefore, extraction frameworks or complex toolchains are necessary. For this purpose \citeauthor{Schmicker.2019}~\cite{Schmicker.2019} propose AndroParse which is written in Golang and uses AAPT and the RAPID library to extract the package name, package version, MD5, SHA1, SHA256, date extracted, file size, permissions, APIs, strings and intents\footnote{The source code can be found at https://github.com/rschmicker/AndroParse/wiki/Querying-the-REST-API. The authors claim to have used this tool to create a dataset of 67,703 harmless and 46,683 malicious APKs, which should be available via API at the address https://64.251.61.74/all. Unfortunately, the address is no longer available.}. Another illustration of a tool chain for analyzing Android applications can be found in \cref{fig:Fig_3_Feature_Extraction_Toolchain_Linares_Vaquez_2016} and was introduced by \citeauthor{LinaresVasquez.2016}~\cite{LinaresVasquez.2016}.

The examined tools are divided into "APK Decompiler and APK Analysis Tools", "Java Analysis Tools", "Hash Tools", and "Helper Tools". Under "APK Decompiler and APK Analysis Tools", all tools necessary to decompile Android APK and those that can directly handle Android APK without intermediate steps were listed. Furthermore, this area contains complex tools which can directly derive deeper information like graphs from the Java code. The "Java Analysis Tools" include all tools that can be used after translating an Android application into pure Java code. "Hash Tools" can generate hash values for comparing purposes or malware fingerprinting. Under the "Helper Tools", all smaller tools that are not directly related to the other topics are listed.

Several problems related to the investigated tools and toolchains were detected during the research. Most of the tools used in the literature cannot keep pace with the development cycles of Android and are not regularly or never maintained. Therefore, they are often not adapted to the actual Android version. For example, Androguard’s last official release is version 3.3.5, released on 18 Feb 2019. At this date, Android Pie (Version number 9, API Level 28) was the current version. Today (September 2022), Android 13 is on the market (Version number 13, API Level 33). This discrepancy leads to the issue that some parts of Android packages that were changed or introduced with newer versions cannot be analyzed entirely. This blind spot leads to undiscovered malware or unrecognized clones. Another major problem with much of the literature is that the underlying systems are often not publicly available, no longer accessible, or not open source. Additionally, the description of how the system’s toolchains are usually inadequate or incomplete, which makes replication difficult. Therefore, most approaches cannot be exactly replicated, and the solutions cannot be verified or compared. For example, a toolchain is described in \citeauthor{LinaresVasquez.2016}~\cite{LinaresVasquez.2016}\footnote{Their system is publicly available under https://www.cs.wm.edu/semeru/data/clandroid/.}. If the authors had not made their system publicly available, replicating their system would not have been possible because the configurations of the underlying tools are not well defined in their paper. Another example is "AndroParse" \cite{Schmicker.2019}, a project written in Golang \cite{go.dev.20240424} that extracts various features using AAPT and the RAPID library. While the source code is still publicly available, the resulting data is unfortunately no longer accessible \footnote{The source code is available at https://github.com/rschmicker/AndroParse/wiki/Querying-the-REST-API. The dataset, which consists of 67,703 harmless and 46,683 malicious APKs, should be available at https://64.251.61.74/all. However, the address is no longer accessible.}.
\begin{table}[htbp]
    \centering
    \caption{Feature extraction tools}
    \label{tab:Tab_3_Feature_extraction_tools}
    \begin{tabular}{lll}
        \toprule
        \textbf{Name} & \textbf{Paper} & \textbf{Category} \\
        
        \midrule
        
        python \cite{Python.org.20240404} & \cite{Chen.2014} & Programming language \\
        
        apktool \cite{Tumbleson.2020} & \cite{Anupama.2022, Batyuk.2011b, Gorla.2014, Hosseini.2021, LinaresVasquez.2016, LinaresVasquez.2014, Shao.2014, Sun.2014, Tian.2015b, Wang.2015, Xu.2018, Zhang.2018} & APK decompiler \\
        
        Baksmali \cite{Baksmali.20240404} & \cite{Blasing.2010, Chen.2014, McLaughlin.2017, Shao.2014, Sun.2014, Zhou.2013, Zhou.2012} & APK decompiler \\
        
        Dedexer \cite{Paller.2009} & \cite{Felt.2011, Sanz.2012} & APK decompiler \\

        dexdump \cite{archlinux.06.09.2022} & \cite{Potharaju.2012} & APK decompiler \\
        
        dex2jar \cite{Pan.2016} & \cite{Crussell.2012, Crussell.2013, Crussell.2015, LinaresVasquez.2016, Shao.2014, Tian.2015b, Wang.2015, Yang.2014} & APK decompiler \\
        
        Androguard \cite{Desnos.2018} & \cite{Crussell.2012, Sahs.2012,Wang.2015,Yang.2014,Zhang.2015} & APK analysis tool \\

        A3E \cite{Azim.2013} & \cite{Shao.2014} & APK analysis tool \\

        IDA \cite{Hex-Rays.2022} & \cite{Hosseini.2021} & APK analysis tool \\

        Keytool \cite{Oracle.2022} & \cite{Wang.2015, Zhou.2012} & APK analysis tool \\
        
        Soot \cite{ValleeRai.1999} & \cite{Chen.2013} & APK analysis tool \\

        aapt \cite{AndroidDevelopers.25.08.2020} & \cite{Anupama.2022, Schmicker.2019, Shao.2014} & Helper Tool \\
        
        BCEL Java Library \cite{ApacheSoftwareFoundation..2020} & \cite{Tian.2015b} & Java analysis tool \\
        
        ckjm \cite{Spinellis.2005} & \cite{Tian.2015b} & Java analysis tool \\
        
        JAD Compiler \cite{Kouznetsov.2001} & \cite{LinaresVasquez.2014} & Java analysis tool \\
        
        JADX Compiler \cite{Skylot.2020} & \cite{Hosseini.2021} & Java analysis tool \\
        
        JClassInfo \cite{Anarxia.2004} & \cite{LinaresVasquez.2016} & Java analysis tool \\
        
        JD Core \cite{GitHub.08.09.2022} & \cite{Wang.2015} & Java analysis tool \\
        
        WALA \cite{IBM.2022} & \cite{Crussell.2012,Crussell.2013,Crussell.2015, LinaresVasquez.2014} & Java analysis tool \\
        
        RAPID library \cite{Zhang.2016b} & \cite{Schmicker.2019} & Java analysis tool \\
               
        LSH code \cite{Andoni.2006} & \cite{Crussell.2013, Crussell.2015} & Hash Tool \\
        
        MinHash \cite{Broder.1997, Broder.2000} & \cite{Crussell.2013, Crussell.2015} & Hash Tool \\ 
        
        7zip \cite{Riehm.16.07.2022} & \cite{LinaresVasquez.2016, LinaresVasquez.2014} & Helper Tool \\
       
        \bottomrule
    \end{tabular}
\end{table}
\subsubsection{APK Decompiler and APK Analysis Tools}
\label{sssec:apk_decompiler_analysis_tools}

There are several tools to decompile and analyze Android applications. In the studied literature, decompilers like the APKTool, Dedexer, Dexdump, Dex2Jar, and Baksmali are used. After decompiling the APK, tools like the keytool, soot, or Androguard extract helpful information. 

One of the most used tools to decompile Android applications is Apktool \cite{Tumbleson.2020}. It is a tool to decompress \cite{LinaresVasquez.2016,LinaresVasquez.2014,Sun.2014} and decompile \cite{Anupama.2022,Batyuk.2011b,Gorla.2014,LinaresVasquez.2016,LinaresVasquez.2014,Wang.2015,Xu.2018,Zhang.2018} Android packages. Furthermore, it contains all features of Baksmali \cite{Sun.2014} and utilities for reverse engineering Android applications \cite{Anupama.2022}. It is possible to extract the compiled binary data \cite{Gorla.2014} and human-readable Smali files \cite{Hosseini.2021,Shao.2014,Xu.2018,Zhang.2018}. The Android manifest and the application resources can also be obtained \cite{Shao.2014,Tian.2015b}. However, according to \citeauthor{Shao.2014}~\cite{Shao.2014}, Apktool might fail to extract the information from an application. According to the authors, a solution could be to use aapt and Baksmali to extract resources and Smali code in separate steps. Dedexer \cite{Paller.2009} and Dexdump \cite{archlinux.06.09.2022} are other tools for decompiling Android applications. Dedexer reads DEX data and translates the data into Java bytecode \cite{Sanz.2012}. The authors say it can generate a directory structure with the identified classes. Dexdump is a disassembler included in the Android SDK \cite{archlinux.06.09.2022}. According to \citeauthor{Potharaju.2012}~\cite{Potharaju.2012}, dexdump creates a dump of all classes and methods derived from the DEX file. In contrast to the other decompilers, Dex2Jar \cite{Pan.2016} allows transferring DEX files directly into JAR files \cite{Crussell.2012, Crussell.2013}. These JAR files are zipped representations of the .class files. According to \citeauthor{Crussell.2013}~\cite{Crussell.2013}, dex2jar has a higher success rate than tools converting directly from DEX to Java.

Another option to handle Android applications is to transform data into a human-readable representation of the Dalvik bytecode, for example Smali. A tool to translate an APK in this format is given by Baksmali, which can reverse engineer classes.dex to Smali code \cite{Blasing.2010, Chen.2014, McLaughlin.2017, Zhou.2013, Zhou.2012}. \citeauthor{Zhou.2013}~\cite{Zhou.2013} extend the Baksmali disassembler so that the tool can directly derive program dependency graphs (PDG). According to \citeauthor{Chen.2014}~\cite{Chen.2014}, an advantage of Baksmali is that Smali code can be compiled back to runnable DEX data.

A further resource that can be used is information about the authors. This data can be mined with the keytool, a tool to manage Android’s key and certificate information \cite{Oracle.2022}. It is implemented in the Android SDK and can authenticate Android applications \cite{GoogleDevelopers.07.09.2022}. \citeauthor{Zhou.2012}~\cite{Zhou.2012} and \citeauthor{Wang.2015}~\cite{Wang.2015} use it in reverse to extract authors’ information.

All previously shown tools can only decompile Android applications and require other tools or complex toolchains to extract features. In contrast to this, Androguard, Soot, Interactive DisAssembler (IDA), and Automatic Android App Explorer (A3E) do not require external support. Soot, which was proposed by \citeauthor{ValleeRai.1999}~\cite{ValleeRai.1999}, can be used for generating high-level output data, such as call graphs, sink-to-source analysis, and data flow analysis.  A3E, which was introduced by \citeauthor{Azim.2013}~\cite{Azim.2013}, is a framework specifically designed to analyze Android Apps directly on smartphones during app runtime. IDA \cite{Hex-Rays.2022} is a powerful disassembler and debugger that is not limited to Android. It serves as an important tool for analyzing application code, enabling the analysis and visualization of code and data flow. The literature reviewed in this work primarily employs these tools for graph construction \cite{Chen.2013, Shao.2014, Hosseini.2021}. 

A very powerful tool to process an APK with only one tool is Androguard. According to the Androguard webpage, Androguard combines all the other tools into a single command line interface application \cite{Desno.2018}. The authors call Androguard "the swiss army knife" in the context of Android analysis because it allows decompiling apps and extracting all needed features like used permissions or different graphs. They also mention that Androguard has better exception handling than other tools. Furthermore,  according to \citeauthor{Yang.2014}~\cite{Yang.2014}, the decompiled results of Androguard are more accurate than other tools like dex2Jar or Baksmali.

\subsubsection{Java Analysis Tools}
\label{sssec:java_analysis_tools}

Many Java analyzers can analyze Android applications because Android is based on Java. JD Core, JAD, and JDAX are Java decompilers that can be used when analyzing the application’s Java code \cite{GitHub.08.09.2022}. JD Core and JAD can transform .class files to Java source code \cite{Kouznetsov.2001}. JADX is the GUI implementation of the JAD decompiler and can also be used for analysis purposes \cite{Skylot.2020}.

Tools like the Apache Commons BCEL Java library, Jclassinfo, the Watson Libraries for Analysis, and the Chidamber and Kemerer Java Metrics are useful for analyzing the underlying structures. The Apache Commons BCEL Java library can transform the byte code files into readable JVM instructions, which can be analyzed to identify method invocations indicating dependencies \cite{ApacheSoftwareFoundation..2020}. Jclassinfo reads Java class files and provides information about the class, dependencies, and more \cite{Anarxia.2004}. The Watson Libraries for Analysis (WALA) \cite{IBM.2022} is an open-source tool initially developed by the IBM T.J. Watson Research Centre in 2006, which delivers tools for the static analysis of Java byte code and JavaScript. Next to other possible uses, in the field of app analysis, WALA is often used to generate program dependency graphs (PDG) \cite{Crussell.2012}. According to the author, WALA can construct PDGs for each method in every class of the Android application directly from the jar files. The Chidamber and Kemerer Java Metrics (short ckjm), written by \citeauthor{Spinellis.2005}~\cite{Spinellis.2005}, is a tool to calculate six different metrics for each class of jar files. These six metrics are the weighted methods per class (WMC), the depth of inheritance tree (DIT), the number of children (NOC), the coupling between object classes (CBO), the response for a class (RFC), and the lack of cohesion in methods (LCOM). In addition, it can also calculate afferent couplings (CA) and the number of public methods (NPM).
\subsubsection{Hash Tools}
\label{sssec:hash_tools}

Hashing is often used to detect malware as well as to detect similar applications. According to \citeauthor{Crussell.2015}~\cite{Crussell.2015}, locality-sensitive hashing (LSH) \cite{Andoni.2006} can efficiently find the nearest neighbors in a large number of vectors. Therefore, LSH uses a lot of different hash functions to raise the possibility of a collision when hashing similar vectors. Another instance of LSH is the MinHash algorithm, which originates from Alta Vista \cite{Broder.1997,Broder.2000}. It was developed to find similar web pages and is also based on the theory of LSH. To do that, features are extracted from the compared examinees. Those features generate hash values based on a set of hashing algorithms. 
\subsubsection{Helper Tools}
\label{sssec:helper_tools}

Besides the previously explained tools, helper tools are not specialized in software engineering. These are often used to fulfill base tasks. An example can be given by 7zip, which is a tool to decompress archives \cite{Riehm.16.07.2022}. In Android analysis, it is often used to extract the .class files from the JAR \cite{LinaresVasquez.2016,LinaresVasquez.2014}. A more complex zip tool is AAPT which allows one to view, create, and update zipped packages like ZIP, JAR, and APK \cite{AndroidDevelopers.25.08.2020}.

\section{EXTRACTABLE ANDROID FEATURES}
\label{sec:extractable_android_features}

The study analyzes the features found in the research projects studied and explains how to use and extract them. Only static features are analyzed; features extracted at runtime, such as CPU usage or phone temperature, are not considered. The analysis is divided into features that can be mined from the app store and features that can be extracted from the APK (see \cref{tab:Tab_4_Overview_Analyzed_Features}), which will be shown in the following chapters. In addition, each feature is described in a subchapter, where all the literature that uses that feature is examined. From the app stores, textual features like the app description, technical features like the reported permissions, and metadata like the rating of the app are analyzed. From the APK file, textual features like strings, technical features like permissions, meta information from the manifest, and statistics derived from these features are examined. In addition, within these chapters, a separation is made between literature from the field of malware detection as well as similarity and clone detection to compare approaches that can be used in different domains. This was done because none of the reviewed papers analyze tools and techniques that cover both domains or compare how the tools, features, and techniques are used in the specific domain. Since most of the literature uses more than one feature, some publications appear in more than one chapter. Not every approach is discussed in detail; a detailed description of the approach can already be found in the underlying literature.
\begin{table}[htbp]
    \centering
    \caption{Overview of analyzed features}
    \label{tab:Tab_4_Overview_Analyzed_Features}
    \begin{tabular}{llp{5cm}l}
        \toprule
        \textbf{Source} & \textbf{Type} & \textbf{Used by} & \textbf{Count} \\
        \midrule
        \multirow{3}{*}{App Store features} 
        & Textual Features & \cite{AlSubaihin.2019, AlSubaihin.2016, Berardi.2015, Bhandari.2013, Chen.2015, Crussell.2012, Gorla.2014, Johann.2017, Pandita.2002, Uddin.2020, Yin.2013} & 11 \\
        & App Store Metadata &  \cite{Berardi.2015, Bhandari.2013, Chen.2015, Crussell.2012, Johann.2017, Pandita.2002,Peng.2012,Potharaju.2012,Sanz.2012,Shabtai.2010,Tian.2015b} & 11 \\
        & Technical Features & \cite{Chen.2015, Pandita.2002,Peng.2012,Sanz.2012,Zhou.2018} & 5 \\
        \midrule
        \multirow{3}{*}{APK features}
        & Technical Features & \cite{Abawajy.2021, Allix.2016, Anupama.2022, Batyuk.2011c,Blasing.2010,Chen.2014,Chen.2013,Crussell.2012,Crussell.2013,Crussell.2015,Felt.2011,Gonzalez.2016,Gonzalez.2015,Gorla.2014,Hamednai.2019,Hanna.2013,Hosseini.2021,Jerome.2014,LinaresVasquez.2016, LinaresVasquez.2014,McLaughlin.2017,Potharaju.2012,Sahs.2012,Sanz.2012,Schmicker.2019,Schmidt.2009c,Shabtai.2010,Shao.2014,Sun.2014,Tian.2015b,Wang.2015,Xu.2018,Xu.2019,Yang.2014,Zhang.2015,Zhang.2018,Zhou.2013,Zhou.2012,Zhou.2018} & 39 \\
        & Meta-information & \cite{Anupama.2022, Chen.2013, Gonzalez.2015, Hamednai.2019,LinaresVasquez.2016,Schmicker.2019, Shabtai.2010,Shao.2014,Sun.2014,Tian.2015b,Zhou.2013,Zhou.2012,Zhou.2018} & 13 \\
        & Textual Features & \cite{Hamednai.2019, Hanna.2013,Sanz.2012,Shabtai.2010,Tian.2015b} & 5 \\
        \bottomrule
    \end{tabular}
\end{table}
\subsection{App Store features}
\label{ssec:app_store_features}

Play Store or other app store features can be collected by crawling the underlying websites. According to \citeauthor{Feizollah.2015}~\cite{Feizollah.2015}, only 3\% of their researched literature uses those features. They refer to data visible in the Play Store before downloading and installing the APK "applications metadata". Furthermore, those features can be used without having or processing the APK. The advantages of these features are that they are easily accessed and collected. It is essential to know that most of these features are registered by the app’s developers, except for statistical information from the app stores. According to \citeauthor{Pandita.2002}~\cite{Pandita.2002}, it is impossible to be sure that the developer does not manipulate this data, e.g., to reach higher downloads. This uncertainty can be described as the most significant disadvantage of this easily accessed data.

\subsubsection{Description}
\label{sssec:description}

In Play Store, each app can be portrayed by a short description with a maximum of 80 words \cite{Google.30.09.2022} and a long description with a maximum of 4000 words \cite{AppRadar.2023}. According to \citeauthor{Berardi.2015}~\cite[586]{Berardi.2015}, "[t]his field briefly presents the app, and it usually lists its characteristics and functionality". They also mention that the description "gives (...) very detailed information about the aim and functionality of the application" \cite[588]{Berardi.2015}. These characteristics lead to a lot of usage across different domains.

\citeauthor{Crussell.2012}~\cite{Crussell.2012},  \citeauthor{Gorla.2014}~\cite{Gorla.2014}, and  \citeauthor{Pandita.2002}~\cite{Pandita.2002} use the description for security purposes. While WHYPER \cite{Pandita.2002} can detect relationships between permissions and the description, CHABADA \cite{Gorla.2014} analyzes relationships between used API calls and the description. WHYPER uses natural language processing (NLP) to explore the description of every single app. Instead of using NLP techniques, CHABADA is clustering the applications based on their app description. Then each cluster’s API calls are analyzed to detect outliers. Clustering is also used by DNADroid \cite{Crussell.2012} as a first indicator to detect cloned applications.

Another area of using the description is to detect similar applications. This approach was followed by \citeauthor{AlSubaihin.2019}~\cite{AlSubaihin.2019}, \citeauthor{AlSubaihin.2016}~\cite{AlSubaihin.2016}, \citeauthor{Berardi.2015}~\cite{Berardi.2015}, \citeauthor{Bhandari.2013}~\cite{Bhandari.2013}, and \citeauthor{Chen.2015}~\cite{Chen.2015}. \citeauthor{Bhandari.2013}~\cite{Bhandari.2013} use the Term Frequency Inverse Document Frequency (TF-IDF) to calculate the cosine distance based on the app description and other metadata from the Play Store. The distance is used to explain the similarity between applications. Building on that, SimApp \cite{Chen.2015} uses Latent Dirichlet Allocation (LDA) to detect high-level app similarities. \citeauthor{AlSubaihin.2016}~\cite{AlSubaihin.2016} identify functionally similar apps with a two-stage clustering. First, they extract features from apps’ descriptions based on \citeauthor{Harman.2012}~\cite{Harman.2012}\footnote{In \cite{Harman.2012}, the authors analyze app relationships from Blackberry app store based on technical-, customer- and business-related features. Their analysis does not use any source code. Only features which can be crawled from the store itself are used.}. In this stage, features from the description are extracted using the Natural Language Toolkit (NLTK). The second stage uses this information to detect relationships between the apps. Four years later \citeauthor{AlSubaihin.2019}~\cite{AlSubaihin.2019} reviewed different app similarity measurement techniques based on textual features and compared them with commonly known approaches. They compared Latent Dirichlet Allocation, Feature Vector Space Model, Vector Space Model + Latent Semantic Analysis, feature vector space model based on collocation-based feature, and feature vector space model based on Dependency-Based Feature.\footnote{Further information and sources about the used models can be found in \cite{AlSubaihin.2019}, chapter 2.}

A special field of similarity detection is detecting functional coverage between applications. Systems that can discover apps with similar functions can also be called app recommendation systems. The advantage of this technique is that apps with similar features can replace dangerous applications with harmless alternatives without limiting functional usage. A first approach was made by \citeauthor{Yin.2013}~\cite{Yin.2013}. First, they calculate the functional coverage between different apps by Latent Dirichlet Allocation (LDA) based on the app description. Then, the authors calculate possible candidates to replace installed apps with an Actual-Tempting model combined with content-based recommendation systems and collaborative filters. Another system that can deliver functionally similar apps was proposed by \citeauthor{Johann.2017}~\cite{Johann.2017} called SAFE (Simple Approach for Feature Extraction). The authors analyze the speech of the description in a deductive human way. Therefore, they implemented "18 part-of-speech  and 5 sentence patterns" \cite[21]{Johann.2017} that the developers frequently followed in the app description. After pre-processing the app description with well-known techniques like removing stop words and tokenization, the authors use the 18 different POS patterns to tag the words in sentences based on linguistic building blocks like "noun-noun" or "verb-noun". After that, the sentences are analyzed by five different patterns, e.g., "send and receive images, videos and sticker" is interpreted as "send images", "send videos", and so on \cite[22]{Johann.2017}. This information is used to compare applications. Furthermore, \citeauthor{Uddin.2020}~\cite{Uddin.2020} compare text-based semantic similarity scoring systems based on descriptions to detect feature similarities of Android applications. In their first approach, they analyze the complete description only pre-processed by well-known text analysis techniques like stop word removal and tokenization. The outcome is analyzed with WordNET \cite{Miller.1995} and Word2Vect \cite{Rehurek.2010}. In their second approach, they extract features of applications from the description with the SAFE \cite{Johann.2017} system. They conclude that the feature-based approach leads to better results.

\subsubsection{Permissions and features}
\label{sssec:permissions_and_features}

It is also possible to get information about used permissions and features from Play Store. This data can be used to detect malicious or vulnerable applications as well as to detect app similarities.

The usage of detecting security leaks was shown by \citeauthor{Peng.2012}~\cite{Peng.2012} and \citeauthor{Zhou.2018}~\cite{Zhou.2018}. According to \citeauthor{Peng.2012}~\cite[241]{Peng.2012}, "[o]ne  of Android's main defense mechanisms against malicious apps is a risk communication mechanism which warns the user about permissions an app requires before being installed, trusting that the user will make the right decision". Nevertheless, this security policy is inefficient because most applications request many permissions so that users do not check their relevance. Therefore, the authors propose a risk score based on Android permissions. In contrast, \citeauthor{Zhou.2018}~\cite{Zhou.2018} use permissions as a first filter stage to detect malicious applications. Therefore, permission-based heuristics for known malware are used to detect possible candidates for further investigation. 

\citeauthor{Chen.2015}~\cite{Chen.2015} and \citeauthor{Sanz.2012}~\cite{Sanz.2012} use permissions and features to detect similar applications. \citeauthor{Sanz.2012}~\cite{Sanz.2012} crawl permissions and features mentioned on Play Store as well as permissions and features found in the manifest. This information is used to categorize Android applications. In contrast, \citeauthor{Chen.2015}~\cite{Chen.2015} only use permissions gathered from Play Store. They transfer them to a Bag of Words and use the (Term Frequency Inverse Document Frequency) TF-IDF. Then, the matrix similarity is computed by the Radial basis function (RBF) kernel. 
\subsubsection{App Rating Score and App Rating counter}
\label{sssec:app_rating_score}

Google and most other app stores use a rating system based on a Likert Scale with a maximum of 5 (1 worst – 5 best). In most cases, the stores provide the average rating of the app given by users and the number of users that rated the app. 

Contrary to the other features, app rating information is usually not used to detect malicious applications because they give no direct information about their security weaknesses. 

\citeauthor{Potharaju.2012}~\cite{Potharaju.2012} propose a system to detect plagiarized applications using app rating as an indicator. Furthermore, rating information is used by \citeauthor{Berardi.2015}~\cite{Berardi.2015}, \citeauthor{Chen.2015}~\cite{Chen.2015}, \citeauthor{Sanz.2012}~\cite{Sanz.2012}, and \citeauthor{Tian.2015b}~\cite{Tian.2015b} to detect similarities in applications. Their base statement is that besides some other features, apps with the same rating could be similar because their rating provides information about the app’s content quality \cite{Chen.2015}. While \citeauthor{Berardi.2015}~\cite{Berardi.2015}, \citeauthor{Chen.2015}~\cite{Chen.2015}, and \citeauthor{Sanz.2012}~\cite{Sanz.2012} use this information for app categorization or classification, \citeauthor{Tian.2015b}~\cite{Tian.2015b} discovered characteristics of high-ranked apps based on the app’s ranking. In their case, the rating is the target value of their system.
%\footnote{This publication is not dedicated to similarity or malware detection. Nevertheless, it is considered here because the authors focus on relevant features that can be used in the analyzed domains. This approach will be continued in the following chapters.}
\subsubsection{Application size}
\label{sssec:app_size}

Application size can also be used as a feature during app analysis to detect similar applications. \citeauthor{Chen.2015}~\cite{Chen.2015} state that two apps with similar sizes could be more related than apps differing in size. This feature is also used in the previously described papers of \citeauthor{Berardi.2015}~\cite{Berardi.2015}, \citeauthor{Potharaju.2012}~\cite{Potharaju.2012}, \citeauthor{Sanz.2012}~\cite{Sanz.2012}, and \citeauthor{Tian.2015b}~\cite{Tian.2015b}. Furthermore, the application size is among the top 20 features in the approach by \citeauthor{Shabtai.2010}~\cite{Shabtai.2010}. During their work, apps within the categories "games" and "tools" are classified using only static features.
\subsubsection{Category}
\label{sssec:app_category}

\citeauthor{Berardi.2015}~\cite{Berardi.2015}, \citeauthor{Peng.2012}~\cite{Peng.2012}, \citeauthor{Potharaju.2012}~\cite{Potharaju.2012} as well as \citeauthor{Tian.2015b}~\cite{Tian.2015b} also use the Play Store categories as features in their approaches. Other work like \citeauthor{Shabtai.2010}~\cite{Shabtai.2010} also uses categories, but not as features. Instead, they use this information as the target vector of their approaches. In \citeauthor{Chen.2015}~\cite{Chen.2015}, besides other information, the app category is used as a feature to detect similar apps. Therefore, apps get clustered according to their categorization because apps in the same category have a high possibility of showing similar functionalities.
\subsubsection{Meta information}
\label{sssec:app_meta_information}

All other features that can be extracted from app stores are described as meta information. First, all features and all authors will be described. Next, an explanation is given of how the authors use those features.

The most mined meta information is the app’s name used by \citeauthor{Bhandari.2013}~\cite{Bhandari.2013}, \citeauthor{Chen.2015}~\cite{Chen.2015}, \citeauthor{Crussell.2012}~\cite{Crussell.2012}, and \citeauthor{Potharaju.2012}~\cite{Potharaju.2012}. The developers’ names and the user reviews are the second most used meta information. The developers’ name is used by \citeauthor{Chen.2015}~\cite{Chen.2015}, \citeauthor{Crussell.2012}~\cite{Crussell.2012}, and \citeauthor{Peng.2012}~\cite{Peng.2012}, and user reviews are used by \citeauthor{Bhandari.2013}~\cite{Bhandari.2013}, \citeauthor{Chen.2015}~\cite{Chen.2015} and \citeauthor{Johann.2017}~\cite{Johann.2017}. In addition to the previously mentioned metadata, \citeauthor{Crussell.2012}~\cite{Crussell.2012} use package and market names. Also, \citeauthor{Potharaju.2012}~\cite{Potharaju.2012} use additional metadata like the update date, the app’s version, the number of app installations, and the application's price. Furthermore, \citeauthor{Chen.2015}~\cite{Chen.2015}use the update text and promotional images.

\citeauthor{Crussell.2012}~\cite{Crussell.2012} use the proposed metadata to detect possible cloned applications. Therefore, these data are used as a first filtering stage by clustering apps with similar meta information. \citeauthor{Peng.2012}~\cite{Peng.2012} use the app names and permissions to clean their dataset from apps generated by app generation tools running with the default settings. \citeauthor{Potharaju.2012}~\cite{Potharaju.2012} use the metadata to detect plagiarized applications. In \citeauthor{Bhandari.2013}~\cite{Bhandari.2013}, all meta information is used to construct the Term Frequency Inverse Document Frequency (TF-IDF). Then those matrices are used to calculate the cosine distance and get a similarity index. \citeauthor{Chen.2015}~\cite{Chen.2015} and \citeauthor{Johann.2017}~\cite{Johann.2017} use that metadata to detect functionally similar applications.

\subsection{APK Features}
\label{ssec:apk_features}

In contrast to features extracted from app stores, data from the APK itself can only be extracted if the app’s APK is available. A more complex toolchain is necessary to extract this data. In \citeauthor{Feizollah.2015}~\cite{Feizollah.2015}, features extracted from APK are called statistic features. The authors also found that 45\% of their analyzed research uses those features among 36\% of papers using Android permissions.

\subsubsection{Code}
\label{sssec:app_code}

The application's source code is one of the most popular sources to detect malware or app similarity. \cref{tab:Tab_5_APK_Alalyze_Techniques} lists the most used and most important techniques to analyze source code in the fields of malware and similarity detection\footnote{None of the analyzed papers cover both fields.}. This table will be followed by a detailed description of how those approaches are used in the scientific context.

\begin{table}[htbp]
    \centering
    \caption{Techniques to analyze source code}
    \label{tab:Tab_5_APK_Alalyze_Techniques}
    \begin{tabular}{lll}
        \toprule
        \textbf{Analysis Techniques} & \textbf{Malware Detection} & \textbf{Similarity Detection} \\
        \midrule
        Call graph (CG) &  \cite{Hosseini.2021, Xu.2018, Xu.2019, Yang.2014,Zhou.2018} &  \cite{Zhang.2015} \\
        Dependency Graph (DG) & \cite{Yang.2014,Zhou.2013} &  \cite{Crussell.2012, Crussell.2013, Crussell.2015} \\
        Abstract Syntax Tree & &  \cite{Potharaju.2012} \\
        Control Flow Graph (CFG) & \cite{Allix.2016, Sahs.2012,Xu.2018,Xu.2019} & \cite{Chen.2014, Sun.2014} \\
        Data Flow Graph (DFG) & \cite{Xu.2018, Xu.2019} & \\
        Control and Data Flow Graph (CDFG) & \cite{Xu.2018, Xu.2019} & \\
        Sequence Analysis (SA) e.g., n-grams excluding graphs & \cite{Anupama.2022, Canfora.2015, Jerome.2014,McLaughlin.2017,Xu.2019,Zhang.2018} & \cite{Gonzalez.2015, Hanna.2013, Wang.2015, Zhou.2012} \\
        Frequency & \cite{Canfora.2015,Xu.2019,Zhang.2018} & \cite{Wang.2015} \\
        Hash & & \cite{Crussell.2013, Crussell.2015, Gonzalez.2016, Hanna.2013, LinaresVasquez.2014,Zhou.2012} \\
        Filter only code & \cite{Gonzalez.2016, Xu.2019, Zhang.2015,Zhou.2018} & \cite{Crussell.2012} \\
        Other data from the code &  \cite{Batyuk.2011b, Shabtai.2010} & \\
        \bottomrule
    \end{tabular}
\end{table}

One of the most important techniques to detect malicious or similar applications are graphs derived from applications’ code. One example of graphs are call graphs which are used by \citeauthor{Hosseini.2021}~\cite{Hosseini.2021}, \citeauthor{Xu.2018}~\cite{Xu.2018}, \citeauthor{Yang.2014}~\cite{Yang.2014}, and \citeauthor{Zhou.2018}~\cite{Zhou.2018} to detect malicious applications as well as by \citeauthor{Zhang.2015}~\cite{Zhang.2015} to detect app similarity. While \citeauthor{Zhou.2018}~\cite{Zhou.2018} use API call graphs, \citeauthor{Hosseini.2021}~\cite{Hosseini.2021}, \citeauthor{Xu.2018}~\cite{Xu.2018}, and \citeauthor{Yang.2014}~\cite{Yang.2014} use function call graphs. Another special kind of call graph was suggested by \citeauthor{Hosseini.2021}~\cite{Hosseini.2021}. They propose a call graph based on native library calls. According to \citeauthor{Hosseini.2021}~\cite{Hosseini.2021}, it is possible to integrate C or C++ code into Android applications with the Android native development kit (NDK). This code is directly transferred into machine code and stored in the lib.so files. This data can be transferred into a call graph with a tool called IDA. Those call graphs are used to run a CNN-LSTM network which can decide which app is benign and which includes malware. On the other hand, \citeauthor{Zhang.2015}~\cite{Zhang.2015} are constructing function call graphs to detect app similarity. Then they detect the main module by finding activities that invoke their decoration APIs for user interaction. After finding the main module, the authors calculate the relations between the modules.

Another type of often-used graph are graphs representing different kinds of dependencies inside the applications. Those graphs are used by \citeauthor{Yang.2014}~\cite{Yang.2014} and \citeauthor{Zhou.2013}~\cite{Zhou.2013} to detect malicious applications and by \citeauthor{Crussell.2012}~\cite{Crussell.2012}, \citeauthor{Crussell.2013}~\cite{Crussell.2013}, and \citeauthor{Crussell.2015}~\cite{Crussell.2015} to detect app similarity. \citeauthor{Zhou.2013}~\cite{Zhou.2013} use a class-based program dependency graph where classes are the nodes and the connections are the relationships between the classes to detect piggybacked applications. They are using the circumstance that cloned apps share the same primary functions. To differentiate between primary and non-primary modules, each connection in the graph is represented by a unidirectional weight. This weight shows the connection strength regarding the relationships like "class inheritance, package homogeneity (...), method calls, and member field references" \cite[186]{Zhou.2013}. These weights are used to merge and cluster the modules. The primary modules can be detected by analyzing the Android manifest and searching the corresponding entry point ACTION.MAIN, as well as comparing the model against references to the main UI or the package name against the application’s name. Then the authors extract a fingerprint-based on "the requested permissions, the Android API calls used, involved intent types (...), the use of native code or external classes, as well as the authorship information" \cite[187]{Zhou.2013}. These features are represented in a vector which can then be compared with the Jaccard distance. In contrast, DrodiMiner \cite{Yang.2014} extracts two-tier behavior graphs based on known malware families. These behavior graphs detect malware in applications and can be derived from a function call graph. The upper layer is a component dependency graph (CDG) representing interactions between activities, services, broadcast receivers, and content providers. The lower layer is a component behavior graph (CBG) which describes the upper layer's functionalities and logic with their underlying API calls. \citeauthor{Crussell.2012}~\cite{Crussell.2012} compare the applications’ program dependency graphs to detect cloned applications. Their work was expanded by \citeauthor{Crussell.2013}~\cite{Crussell.2013} and \citeauthor{Crussell.2015}~\cite{Crussell.2015}. In this work, the authors propose AnDarwin, a system to detect similarities. AnDarwin constructs a program dependency graph (PDG) based on the data flow for each application. This Data PDG represents the relationship between data inside a function and can be used to detect similar code blocks between applications. Another approach to measuring similarities in applications was proposed by \citeauthor{Chen.2014}~\cite{Chen.2014}. 

One of the most basic graphs are abstract syntax trees (AST). This graph builds the base for a lot of others. The elementary AST was used by \citeauthor{Potharaju.2012}~\cite{Potharaju.2012}. To extract the AST of an APK, the authors use the built-in Android disassembler dexdump to generate a dump file including all classes and methods. To achieve their goals, they have modified the dexdump functions dumpClass, dumpMethod, dumpCode, dumpSField, and dumpIField. The outcome is used to generate an abstract syntax tree (AST) which can be used to construct a fingerprint of each method. The fingerprint contains information about how many virtual and direct functions are called by a method, how many arguments and parameters are passed through a method, and which sub-methods are called. This information is used to detect plagiarized applications with three different approaches, the symbol-coverage, the AST-distance, and the AST-coverage. The symbol coverage can only be used if the code is not obfuscated. In this case, "[t]he coverage of an application A\textsubscript{i} by A is computed as the number of classes and methods in A\textsubscript{i} that also exist in A divided by the total number of classes and methods in A\textsubscript{i}" \cite[113]{Potharaju.2012}. If the code is level-1 obfuscated, the AST-distance calculates the Euclidian distance between the previously shown fingerprints to measure app similarity. Level-2 obfuscation can be handled with the AST-coverage approach. Therefore, the symbol-coverage approach and the AST-distance will be combined.

A specialization of an AST is a control flow graph (CFG), which represents the application's control flow (CFG). According to \citeauthor{Xu.2018}~\cite[178]{Xu.2018}, CFGs reflect "what a program intends to behave (e.g., opcodes) as well as how it behaves (e.g., possible execution paths), such that malware behaviour patterns can be captured easily by this feature". These graphs are used by \citeauthor{Allix.2016}~\cite{Allix.2016}, \citeauthor{Sahs.2012}~\cite{Sahs.2012}, \citeauthor{Xu.2018}~\cite{Xu.2018}, and \citeauthor{Xu.2019}~\cite{Xu.2019} to detect malware. It was also used in the field of detecting similar apps by \citeauthor{Chen.2014}~\cite{Chen.2014} and \citeauthor{Sun.2014}~\cite{Sun.2014}. \citeauthor{Sahs.2012}~\cite{Sahs.2012} build a Control Flow Graph for every method found in the manifest file to detect malware. To avoid metamorphic techniques that are often used to suppress malware detector findings, the authors rewrite parts of the CFG by merging "consecutive instruction blocks", "unconditional jumps with the instruction block they jump to", and "consecutive conditional jumps" \cite[142]{Sahs.2012}. CDGDroid \cite{Xu.2018} also uses a control flow graph (CFG) to detect malicious applications by analyzing the control flow in malicious applications and describing control flows typical for malware. In \citeauthor{Xu.2019}~\cite{Xu.2019}, the authors expand CDGDroid to automatically classify and characterize Android malware families. Therefore, the graphs get weighted, and the API calls are combined with their corresponding permissions. An overview of different approaches to using CFGs is given by \citeauthor{Allix.2016}~\cite{Allix.2016}. Next to some other graphs, the authors test CFG-based techniques to detect malicious applications. The CFGs are derived from the extracted Dalvik bytecode code. This code level contains the most information from the original Java byte code. Each CFG is divided into its basic blocks, and each app is represented by its blocks. Then a Boolean feature vector with all basic blocks is built for each app. This is the fingerprint of the application. The authors justify this approach with the fact that the semantics and structure of the software can better be caught with data blocks than keeping low-level information like variable names or registers because that information can easily be changed and is not independent of the program structure. As mentioned before, \citeauthor{Chen.2014}~\cite{Chen.2014} calculate the geometrical centroid of the CFGs. This technique was derived from force calculation in physics and can be used to compare applications based on their centroids. The centroid is based on a 3D-based method control flow graph (3D-CFG), where "x is the sequence number in the CFG. y is the number of outgoing edges of the node. z is the depth of loop of the node" \cite[178]{Chen.2014}. Also, \citeauthor{Sun.2014}~\cite{Sun.2014} use CFGs to detect app similarity. As mentioned in chapter \ref{sssec:api_calls}, they are used to construct component-based control flow graphs.

Another specialization of the AST is data flow graphs (DFGs). According to \citeauthor{Xu.2018}~\cite{Xu.2018}, DFGs represent the interactions between variables and operations and define how variables are processed during the program steps. DFGs are used by \citeauthor{Xu.2018}~\cite{Xu.2018} and \citeauthor{Xu.2019}~\cite{Xu.2019} for malware detection. Therefore, the graphs are transferred into matrices which will then be analyzed with a CNN to detect malicious applications. \citeauthor{Xu.2018}~\cite{Xu.2018} and \citeauthor{Xu.2019}~\cite{Xu.2019} also use the combination of CFG and DFG.

Another option to analyze Android applications is to use sequence analysis techniques like n-grams. \citeauthor{Canfora.2015}~\cite{Canfora.2015}, \citeauthor{Jerome.2014}~\cite{Jerome.2014}, \citeauthor{McLaughlin.2017}~\cite{McLaughlin.2017}, \citeauthor{Wang.2015}~\cite{Wang.2015}, \citeauthor{Xu.2018}~\cite{Xu.2018}, and \citeauthor{Zhang.2018}~\cite{Zhang.2018} use sequence-related techniques for malware detection while \citeauthor{Gonzalez.2015}~\cite{Gonzalez.2015}, \citeauthor{Hanna.2013}~\cite{Hanna.2013}, and
\citeauthor{Zhou.2012}~\cite{Zhou.2012} use sequence analysis techniques to detect similar applications. In detail, \citeauthor{Canfora.2015}~\cite{Canfora.2015}, \citeauthor{Jerome.2014}~\cite{Jerome.2014}, \citeauthor{Xu.2018}~\cite{Xu.2018}, and \citeauthor{Zhang.2018}~\cite{Zhang.2018}  explicitly use n-grams. An example of how to use n-grams can be given by analyzing the process of n-gram analysis done by \citeauthor{Jerome.2014}~\cite{Jerome.2014}. To derive n-grams, a window slides above the entire code. After each window step, an n-gram is extracted. Then all operands are removed because they are easy to change \cite{Jerome.2014} or obfuscate \cite{Zhou.2012}. In the end, a feature vector is constructed, which represents the occurrence of each unique n-gram. Instead of deriving n-grams from source code, CDGDroid \cite{Xu.2019} extracts the n-grams from the previously mentioned graphs. Another kind of sequence analysis was proposed by \citeauthor{Wang.2015}~\cite{Wang.2015}. The authors are using a two-stage based approach. The first stage is based on API call sequences. The second fine-grained detection stage is based on Boreas\footnote{Boreas is a counting-based code clone detection approach based on matching variables used in the applications \cite{Yuan.2011,Yuan.2012}.}. \citeauthor{McLaughlin.2017}~\cite{McLaughlin.2017} use a convolutional neural network (CNN) to detect malware. This approach allows them to use a long sequence of concatenated opcode n-grams. After removing the operands, this long sequence is transformed into an on-hot-encoded vector. 

\citeauthor{Zhou.2012}~\cite{Zhou.2012} use sequence analysis techniques to detect app similarity. Therefore, instruction sequences are mined with a sliding window, and the operands are discarded. Instead of removing all operands from the n-grams, \citeauthor{Hanna.2013}~\cite{Hanna.2013} keep the constant data like const-strings, making the further analysis more fine-grained. Furthermore, code containing reflections can only be compared if the strings are kept. \citeauthor{Gonzalez.2015}~\cite{Gonzalez.2015} investigate the usage of opcode and bytecode n-grams to detect app similarity. Together with the previously mentioned metadata, the authors build a two-stage system using n-grams to calculate app distances with the Simplified Profile Intersection (SPI) proposed by \citeauthor{Frantzeskou.2006}~\cite{Frantzeskou.2006}. This design allows the authors to detect multiple levels of relationships between applications such as siblings, twins, step-siblings, and cousins.

A technique that can be used additionally to sequence-based approaches is to count the frequency of occurrence of unique sequences. This technique was used by \citeauthor{Canfora.2015}~\cite{Canfora.2015}, \citeauthor{Xu.2019}~\cite{Xu.2019}, and \citeauthor{Zhang.2018}~\cite{Zhang.2018} to detect malware and by \citeauthor{Wang.2015}~\cite{Wang.2015} to detect app similarities. \citeauthor{Canfora.2015}~\cite{Canfora.2015} and \citeauthor{Zhang.2018}~\cite{Zhang.2018} use the frequency of opcode n-grams to detect malware. They compute statistics between the distance of opcode frequencies in benign and malicious applications. Another way to use opcode frequency was shown by \citeauthor{Xu.2019}~\cite{Xu.2019}. The authors expand their tool CDGDroid to detect malware families based on n-grams frequencies. In the domain of detecting similar applications, \citeauthor{Wang.2015}~\cite{Wang.2015} use the frequency of API call sequences.

Another technique that can be used next to the ones previously shown is calculating hashes and using them to find malicious or similar applications. This technique was used by \citeauthor{Crussell.2012}~\cite{Crussell.2012}, \citeauthor{Crussell.2013}~\cite{Crussell.2013}, \citeauthor{Crussell.2015}~\cite{Crussell.2015} \citeauthor{Hanna.2013}~\cite{Hanna.2013}, \citeauthor{LinaresVasquez.2014}~\cite{LinaresVasquez.2014}, and \citeauthor{Zhou.2012}~\cite{Zhou.2012} and  to detect similar applications. \citeauthor{LinaresVasquez.2014}~\cite{LinaresVasquez.2014} use the hash values of all classes found in the application to compare applications. \citeauthor{Crussell.2013}~\cite{Crussell.2013} and \citeauthor{Crussell.2015}~\cite{Crussell.2015} generate hash values over the semantic vectors derived from the PDG to detect similar applications. Another possibility of using hashes was proposed by \citeauthor{Zhou.2012}~\cite{Zhou.2012}. The authors generated a fuzzy hash over their extracted program instruction sequences to generate an application’s fingerprint. The hash generator cuts the hash generation on a specific reset point to get comparable hash values. By doing this, the code is sliced into smaller pieces, and each piece gets its hash. Ultimately, an overall hash is calculated by concatenating the hash values of the smaller code slices. This hash can compare Android applications by calculating the edit distance between two fingerprints. A similar approach was proposed by \citeauthor{Hanna.2013}~\cite{Hanna.2013} within Juxtapp. Instead of unlimited instruction sequences, they use overlapping n-grams as a starting point for their djb2 hash values. These hashes are transferred into a feature vector where each n-gram is represented by a single bit. Then the Jaccard distance between those feature vectors is used to calculate app similarity. 

Filtering techniques can reduce resource usage during calculation by minimizing the search space. Filtering techniques were used by \citeauthor{Gonzalez.2016}~\cite{Gonzalez.2016}, \citeauthor{Xu.2019}~\cite{Xu.2019}, and \citeauthor{Zhang.2015}~\cite{Zhang.2015} and  to find malicious applications and by \citeauthor{Crussell.2012}~\cite{Crussell.2012} to find app similarities.  \citeauthor{Zhang.2015}~\cite{Zhang.2015} filter all function calls which are not meaningful for detecting malware. Their system keeps graphs only if the FCG parts are loosely coupled and end with an abusable API call. Another approach was proposed by \citeauthor{Gonzalez.2016}~\cite{Gonzalez.2016} by using bloom filters to detect different kinds of reused code. For this purpose, some known sources for code reuse are crawled like Android SDK-related base functions, common (advertisement) libraries, or malware code snippets. Then this code is hashed and a bloom filter is constructed. This filter detects the presence and the kind of reused code, like using a library, implementing ads, or using the app for piggybacking malware. \citeauthor{Xu.2019}~\cite{Xu.2019} also use filtering techniques. They filter the opcode n-grams according to their weight and the frequency of opcode occurrence. 

Filtering techniques are also used to identify similar applications. \citeauthor{Crussell.2012}~\cite{Crussell.2012} filter their graphs by removing parts that are not meaningful to detect cloned applications. For example, code from third-party libraries is removed to make the system more robust.

In addition to graph or code sequence-based approaches, code analyses can extract metadata from the code. \citeauthor{Shabtai.2010}~\cite{Shabtai.2010} use code analysis techniques to extract data like strings, types, classes, prototypes, methods, fields, static values, inheritance, and opcodes. These data are used to classify Android applications. Another approach to using code was shown in \cite{Batyuk.2011b}. In their work, the authors search for code parts violating privacy. Then they develop a method to autopatch those code fragments and generate a patched application.

\subsubsection{API Calls}
\label{sssec:api_calls}

According to Google \citeauthor{GoogleDevelopers.2022}~\cite{GoogleDevelopers.2022}, 4821 Android-related API calls currently exist. Android API calls are essential to building an application because they connect the operating system and the hardware components with the application. Another important factor why API calls are often used is that their modification during API name obfuscation is impossible \cite{Wang.2015}.

In general, API calls can be interpreted in multiple ways. According to \citeauthor{LinaresVasquez.2016}~\cite{LinaresVasquez.2016}, API calls can be separated by their source. One source is the Android framework itself because it provides many (operational) system-related commands. Another source could be functions provided by third-party developers like Unity \cite{UnityTechnologies.2023}. All API calls provided by the developer can be regarded as internal calls. \citeauthor{Gorla.2014}~\cite{Gorla.2014} suggest differentiating API calls into sensitive API calls and regular ones. Sensitive API calls are API calls that are based on user permissions. According to \citeauthor{GoogleDevelopers.2022}~\cite{GoogleDevelopers.2022} and \citeauthor{Zhang.2018}~\cite{Zhang.2018}, API calls can be divided into "package name", possibly "sub package name", "class name", and "function name". E.g., LAndroid/net/ConnectivityManager/getNetworkCapabilities means the function "getNetworkCapabilities" is located in the package "ConnectivityManager" which comes from the package "LAndroid", located in the Subpackage "net" where LAndroid indicates that this is an Android native API Call.

\citeauthor{Blasing.2010}~\cite{Blasing.2010},  \citeauthor{Chen.2013}~\cite{Chen.2013}, \citeauthor{Schmidt.2009c}~\cite{Schmidt.2009c}, \citeauthor{Zhang.2018}~\cite{Zhang.2018}, and \citeauthor{Zhou.2018}~\cite{Zhou.2018}  use API calls to detect malicious application behavior. \citeauthor{Zhou.2018}~\cite{Zhou.2018} investigate Android API calls based on API call graphs. They also monitor the parameters and compare all with previously derived malware heuristics. In their approach, API calls are also used to detect unknown zero-day exploits by monitoring dynamically loaded code by observing the corresponding API call from the class "DexClassLoader". If dynamically loaded code is detected, the interactions between the Java code and the Android API calls are monitored in a live system to investigate malicious behaviours. Building on their previous work,  \citeauthor{Zhou.2013}~\cite{Zhou.2013} use API calls to generate a fingerprint to detect piggybacked applications. \citeauthor{Gorla.2014}~\cite{Gorla.2014} use API calls to find outliers of app behavior. \citeauthor{Chen.2013}~\cite{Chen.2013} developed the PEGASUS system, which can detect malicious behavior based on the "temporal properties of the interaction between an application and the Android event system" \cite[1]{Chen.2013}. A permission event graph is constructed to map the temporal transitions between permissions and lifecycle-related components, including their API call usage based on the related events. \citeauthor{Zhang.2018}~\cite{Zhang.2018} use operand sequences derived from API calls for malware detection. For this purpose, they only keep the API call package names. \citeauthor{Blasing.2010}~\cite{Blasing.2010} and \citeauthor{Schmidt.2009c}~\cite{Schmidt.2009c} use monitoring techniques to analyze system and library calls. They extract function call lists and compare them against malware heuristics.

Another application scenario of using API calls is to detect plagiarism, like finding reused code \cite{Sun.2014} or cloned Android applications \cite{Wang.2015} as well as (functional) similar Android applications \cite{Hamednai.2019,LinaresVasquez.2016}. \citeauthor{Sun.2014}~\cite{Sun.2014} construct a component-based control flow graph (CB-CFG) where API calls are the nodes, and the edges represent the control flow. With this approach, the authors can detect reused code. \citeauthor{Wang.2015}~\cite{Wang.2015} propose a three-stage-based system to detect cloned Android applications. In the first stage, a feature pre-processing is done. In the second, coarse-grained stage, possible clones can be detected by comparing the "call frequencies of different Android APIs" \cite[74]{Wang.2015}. The third stage is described in the chapter "Code". \citeauthor{LinaresVasquez.2016}~\cite{LinaresVasquez.2016} propose a system to detect similar Android applications. In this approach, they are using several features to fingerprint applications. One of these features are API calls. \citeauthor{Hamednai.2019}~\cite{Hamednai.2019} use API calls to detect functionally similar applications because, together with permissions, API calls can capture the app’s behaviors and functionalities. Additionally, the authors capture all Android API calls from the Android developer page, merge them with the "Android.jar" file, and compare those with their extracted ones. Also, the authors divide API calls into API-class and API-method, where API-class only contains the API package name and API-method contains the package name and class name.

Furthermore, some other domains were covered by \citeauthor{Schmicker.2019}~\cite{Schmicker.2019} and \citeauthor{Tian.2015b}~\cite{Tian.2015b}. While \citeauthor{Schmicker.2019}~\cite{Schmicker.2019} use API calls in their own malware-related Android feature extraction framework, \citeauthor{Tian.2015b}~\cite{Tian.2015b} use API calls to detect characteristics of high-rated apps.

\subsubsection{Permissions}
\label{sssec:permissions}

Android’s security is based on a permission system where apps have to request permissions granted or declined by the user \cite{Almomani.2020}. In Android 13 API 33, 206 Android permissions are listed, which can be divided into the protection levels normal-, dangerous- and signature- / system-permissions\footnote{A detailed description of Android permission levels can be found in \cite{Felt.2011}, chapter 2.1 "Android Background".} \cite{AndroidDevelopers.21.06.2022b, GoogleDevelopers.2022}. There are two different ways to set those permissions. One option is to request permissions during installation. The other option is to request permissions by user interaction during runtime \cite{AndroidDevelopers.21.06.2022b}. Permissions granted during installation are typically not dangerous for the user and can be changed later in the Android app options. Those permissions must be declared in the Android manifest file \cite{AndroidDevelopers.21.06.2022}.

According to \citeauthor{Sanz.2012}~\cite{Sanz.2012}, the high involvement of user interaction leads to security risks because their security consciousness is limited, and they fail to read permissions requested before installation. Furthermore, \citeauthor{Almomani.2020}~\cite{Almomani.2020} found that a majority of users agree to the listed permissions without fully understanding them. According to \citeauthor{Felt.2011}~\cite{Felt.2011}, one-third of all Android applications are overprivileged. To avoid this, the authors developed Stowaway, a tool that compares API calls and the underlying permissions. 

For malware detection, permissions are used by \citeauthor{Abawajy.2021}~\cite{Abawajy.2021}, \citeauthor{Anupama.2022}~\cite{Anupama.2022}, \citeauthor{Chen.2013}~\cite{Chen.2013}, \citeauthor{Sahs.2012}~\cite{Sahs.2012}, and \citeauthor{Zhou.2013}~\cite{Zhou.2013}. According to \citeauthor{Abawajy.2021}~\cite{Abawajy.2021}, most malware can be detected by changed or overfitted permissions because malware authors commonly exploit these. In addition, \citeauthor{Abawajy.2021}~\cite{Abawajy.2021} investigate which feature selection techniques could perform best in the area of malware detection. \citeauthor{Anupama.2022}~\cite{Anupama.2022} consider permissions as one of two features in their static analysis part. They use them to generate a binary feature vector (0 absent; 1 present). This feature vector is used to train diverse AI systems to detect malicious applications. \citeauthor{Chen.2013}~\cite{Chen.2013} propose the PEGASUS system to detect malicious apps based on the temporal connection between permissions and API calls. Regarding the ability to generate custom permissions, \citeauthor{Sahs.2012}~\cite{Sahs.2012} further divide them into standard built-in permissions and non-standard permissions. The former are handled by setting a bit in a features vector if permission is required in the application. The latter are split into the segments "prefix (usually "com" or "org"), the organization and product section, and the permission name" \cite[142]{Sahs.2012}.  \citeauthor{Zhou.2013}~\cite{Zhou.2013} use permissions to detect piggybacked apps.

On the other hand, \citeauthor{Hamednai.2019}~\cite{Hamednai.2019}, \citeauthor{LinaresVasquez.2016}~\cite{LinaresVasquez.2016}, \citeauthor{Sanz.2012}~\cite{Sanz.2012}, and \citeauthor{Shabtai.2010}~\cite{Shabtai.2010} compare the permission usage to detect app similarities.  

As shown in the previous chapter, \citeauthor{Schmicker.2019}~\cite{Schmicker.2019} also use permissions in their own malware-related Android feature extraction framework, while \citeauthor{Tian.2015b}~\cite{Tian.2015b} use them to detect characteristics of high-rated apps.

%\citeauthor{LinaresVasquez.2016}~\cite{LinaresVasquez.2016}, \citeauthor{Sahs.2012}~\cite{Sahs.2012}, \citeauthor{Sanz.2012}~\cite{Sanz.2012}, \citeauthor{Schmicker.2019}~\cite{Schmicker.2019}, \citeauthor{Shabtai.2010}~\cite{Shabtai.2010}, \citeauthor{Tian.2015b}~\cite{Tian.2015b}, and \citeauthor{Zhou.2013}~\cite{Zhou.2013}. 
\subsubsection{Metadata from the Manifest}
\label{sssec:manifest_metadata}

Besides permissions, other helpful information can be mined from the manifest file, for example intent filters, providers, and queries. A complete list can be found under \citeauthor{AndroidDevelopers.21.06.2022}~\cite{AndroidDevelopers.21.06.2022}.

\citeauthor{Shabtai.2010}~\cite{Shabtai.2010}, \citeauthor{Shao.2014}~\cite{Shao.2014}, and \citeauthor{Tian.2015b}~\cite{Tian.2015b} use data from the manifest to generate statistics. \citeauthor{Shabtai.2010}~\cite{Shabtai.2010} use elements, attributes, namespaces, strings, and actions. \citeauthor{Shao.2014}~\cite{Shao.2014} use the number of activities, the number of permissions, and the number of intent filters. \citeauthor{Tian.2015b}~\cite{Tian.2015b} use the count of specified services and activities in the manifest as a statistical value to detect high-ranked Android apps. 

\citeauthor{Zhou.2012}~\cite{Zhou.2012} use intent filters to detect repacked applications. \citeauthor{Zhou.2013}~\cite{Zhou.2013} extract the "involved intent types (which represent the way for inter-component or inter-process communication)" \cite[187]{Zhou.2013} for piggybacked app detection. Furthermore, \citeauthor{Zhou.2018}~\cite{Zhou.2018} use services and receivers to detect malicious applications based on malware heuristics.

\citeauthor{Anupama.2022}~\cite{Anupama.2022} and \citeauthor{Chen.2013}~\cite{Chen.2013} extract application components like activities, services, broadcast receivers, and content providers. \citeauthor{Chen.2013}~\cite{Chen.2013} use those to construct relationships between the UI, the API calls, and the temporal colorations between all of them. \citeauthor{Anupama.2022}~\cite{Anupama.2022} build binary feature vectors from those components.

\citeauthor{Hamednai.2019}~\cite{Hamednai.2019} and \citeauthor{LinaresVasquez.2016}~\cite{LinaresVasquez.2016} use data from the manifest to detect app similarities. While \citeauthor{LinaresVasquez.2016}~\cite{LinaresVasquez.2016} only extract intents, \citeauthor{Hamednai.2019}~\cite{Hamednai.2019} collect several features from the manifest to fulfill this purpose. \citeauthor{Hamednai.2019}~\cite{Hamednai.2019} collect hardware and software components required by the app, app components like activities, services, broadcast receivers, content providers, and intent filters. This information is used to construct feature vectors to check app similarity. Furthermore, the package name in the manifest is captured. For this purpose, package names are split into separate components like "weatherforecast" and will be interpreted as "weather" and "forecast".

\subsubsection{UI-related data}
\label{sssec:ui_related_data}

The connection between GUI usage and the expected app behavior can be used to detect malicious applications. \citeauthor{Chen.2013}~\cite{Chen.2013} extract "user interface entities such as buttons and widgets, and their relevant event-handlers" \cite[4]{Chen.2013}. This data is used to check if an app only provides the behavior expected by the user or if an app shows malicious behavior under the surface.

Because the GUI is an essential part of an APK, \citeauthor{Shao.2014}~\cite{Shao.2014} use it to detect cloned applications. The base idea is that resources and code are divided into separate files to make the application simple to maintain. The relationships between those components can then be used to detect repacked Android applications because GUI elements (layouts) and event handlers are mostly not modified by malware authors because the interaction between those elements represents the app's functionality. This mechanism could change the original app’s functionality, making it less likely for the malware to be undetected or spread.

\citeauthor{Taba.2014}~\cite{Taba.2014} and \citeauthor{Tian.2015b}~\cite{Tian.2015b} investigated how user interface complexity relates to the ranking in the Google Play Store. They achieve contradictory results, which may be grounded in the different approaches used to identify the UI complexity. \citeauthor{Taba.2014}~\cite{Taba.2014} based the complexity calculation on UI elements like input and output and clustered them based on textual features like strings extracted from each UI. This data is set concerning the application’s rank in the Play Store to find relations between UI and perceived app quality. On the other hand, \citeauthor{Tian.2015b}~\cite{Tian.2015b} use GUI elements and their occurrence in the resource folders to detect highly rated applications.

\subsubsection{Sensors}
\label{sssec:sensors}

Sensors are all hardware elements that can detect changes in the environment of the device, like GPS position, temperature, or acceleration. \citeauthor{LinaresVasquez.2016}~\cite{LinaresVasquez.2016} utilize the information on sensor usage as a feature to detect app similarity. For this purpose, they have written a sensors extractor using APKs’ zip file as input. In \citeauthor{Batyuk.2011c}~\cite{Batyuk.2011c}, sensor information like geodesic coordinates and ambient loudness is used to obtain information about the current context of phone usage, for example, using the phone in a business context or for private purposes. Dependent on this information, the phone settings can be adjusted according to the security conditions of the context.

\subsubsection{Identifiers}
\label{sssec:identifiers}

In Android, identifiers can be used to track the user’s behavior when using applications. This is primarily used for market analysis purposes. \citeauthor{LinaresVasquez.2016}~\cite{LinaresVasquez.2016} use identifiers as one feature in their work to detect similar applications.

\subsubsection{Printable Strings}
\label{sssec:printable_strings}

Printable strings can be found in many places of an APK, like inside the code, the folder structure, or the underlying files. According to \citeauthor{Shabtai.2010}~\cite{Shabtai.2010}, string features are "meaningful plain-text strings that are encoded in programs files" \cite[330]{Shabtai.2010}. \citeauthor{Sanz.2012}~\cite{Sanz.2012} and \citeauthor{Shabtai.2010}~\cite{Shabtai.2010} extract this information from dex files and count their frequency of occurrence. This information is used to classify Android applications into different categories. \citeauthor{Tian.2015b}~\cite{Tian.2015b} extract strings from the application’s root directory, which includes a "resource" folder containing, e.g., strings. This information is used to detect high-ranked applications. \citeauthor{Hamednai.2019}~\cite{Hamednai.2019} use strings found in dex files and strings found in xml files in "/res/values/" folder to gather app-related string information. This data gets filtered to remove duplicates, for example, caused by double values against other features like API calls. \citeauthor{Schmicker.2019}~\cite{Schmicker.2019} use the AAPT tool to extract strings in their Android feature extraction framework.

\subsubsection{Statistical features}
\label{sssec:statistical_features}

Statistical Android features are all kinds of data that represent some numerical aspect of the application, like the size of the application or the count of files included in the APK. 

\citeauthor{Shabtai.2010}~\cite{Shabtai.2010} explore statistical features in the APK and the manifest. From the APK, they extract ".apk size, number of zip entries, number of files for each file type, common folders and more" \cite[331]{Shabtai.2010}. From the xml, they mine the "[c]ount of xml elements, attributes, namespaces, and distinct strings" \cite[331]{Shabtai.2010}. They also collect features for "each element name (e.g., count of distinct actions, distinct categories), [f]eatures for each attribute name [and] [f]eatures for each attribute type (e.g., number, string)" \cite[331]{Shabtai.2010}.

\citeauthor{Zhou.2018}~\cite{Zhou.2018} use the archive structure to detect malware based on malware heuristics. For this purpose, the archive is decompressed, and the internal structure is monitored.

\citeauthor{Gonzalez.2015}~\cite{Gonzalez.2015},  
\citeauthor{Schmicker.2019}~\cite{Schmicker.2019}, and \citeauthor{Shao.2014}~\cite{Shao.2014} use statistics derived from different data types. While \citeauthor{Shao.2014}~\cite{Shao.2014} use the average number of .png files and .xml files per directory, \citeauthor{Gonzalez.2015}~\cite{Gonzalez.2015} use statistics derived from the number as well as the (average) size of .apk, .zip, .java, .jar, images, libraries, and binary files. \citeauthor{Schmicker.2019}~\cite{Schmicker.2019} use the file size as well as the date.
\subsubsection{Metadata}
\label{sssec:metadata}

Android-related metadata like the inner structure, security policies, organization, and ownership characterize the application. \citeauthor{Sun.2014}~\cite{Sun.2014} extract some information from the META-INF file, like detailed information about the application’s author and certificate file cert.rsa. Furthermore, \citeauthor{Gonzalez.2015}~\cite[441]{Gonzalez.2015} use the "certificate serial number, the hash value (md5) for the .apk container and a .dex file (md5), and a list of internal files’ names with corresponding hash values (md5)" to detect app similarities. Additionally, \citeauthor{Schmicker.2019}~\cite{Schmicker.2019} extract several hash values like MD5, SHA1 and SHA256.

\section{CREATING A UNIVERSAL, INTERDISCIPLINARY DATASET}
\label{sec:creating_universal_interdisciplinary_dataset}

%Einleitung
After describing the data sources, extraction tools, and features most often used to analyze research questions in similarity, clone, and malware detection, this section shifts its focus to defining the requirements for a multipurpose dataset. First, the necessity of such a dataset is presented from the scientific and user perspective. Second, the challenges of existing approaches identified in the reviewed literature are highlighted. Based on this, guidelines and the schematic process for generating the dataset are described.

%Notwendigkeit aus wissenschaftlicher Sicht
From a scientific point of view, creating a universal dataset fosters a research environment that enables validation and replication, facilitating and accelerating the development of new methods and models. Furthermore, a universal dataset can be applied not only to specific use cases\footnote{Researchers typically extract only the data relevant to their specific research, making comparisons to alternative approaches, which typically require different features, difficult.}, but to a variety of other problems, leading to more general, consistent, and robust solutions and enabling new fields of research. Additionally, an existing dataset significantly reduces the time and effort involved in data collection. By consolidating all relevant features in one dataset, researchers can work more efficiently and at the same time exploit synergy effects between malware, clone, and similarity analysis. For instance, a system for detecting clones could be expanded to detect subtle differences in potentially identical applications, which can offer a way to detect (previously unknown) malicious code fragments \cite{Zhou.2013}.\footnote{A system to detect piggybacked apps was proposed by \citeauthor{Zhou.2013}~\cite{Zhou.2013}. However, the authors did not make their dataset publicly available.} This opens up new possibilities by enabling collaboration and knowledge sharing between different research groups and disciplines.

%Notwendigkeit aus nutzenden Sicht
In addition to the benefits to the research community, future systems based on a universal dataset promise improved security and usability for users. These systems are not only able to detect malware but also offer specific recommendations for action, such as alternatives to potentially dangerous applications.\footnote{For example, \citeauthor{LinaresVasquez.2016}~\cite{LinaresVasquez.2016} proposed a system for detecting similar Android applications that could be used for this purpose.} In addition, they could also implement recommender systems\footnote{For instance, systems that proactively show apps using more permissions than the app functionality requires. An example was provided by \citeauthor{Gorla.2014}~\cite{Gorla.2014}. However, they only analyzed which permissions are necessary based on the app's description but not on its functionality.}, which can further reduce the burden on users. These points are particularly important because user studies suggest that risk mitigation systems only achieve their full potential when the perceived cost of risk mitigation does not exceed the perceived effectiveness \cite{Alsaleh.2017, Hansen.2004, Liang.2009, Sasse.2005}. Providing security-relevant information to technically-savvy users can overwhelm non-technically savvy users, which in turn represents a security risk for everyone. Therefore, different user groups must be addressed through adapted security mechanisms and options for action \cite{Hansen.2004, Sasse.2005}. Systems addressing the individual needs of users therefore enable an improved user experience for all user groups, regardless of their technical background.

% Zusammenfassung der Notwendigkeit
This paper identified numerous challenges associated with the creation of a universal dataset. A significant obstacle to the research, validation, and replication of current and future threat protection systems are unpublished or no longer available datasets. In addition, the reviewed literature suffers from deficiencies in data quality, data collection, data storage, and documentation.\footnote{As described in section \ref{ssec:gather_andorid_apps}, many authors do not provide clearly defined data collection methods or make datasets publicly available. Furthermore, some authors publish only a description of the dataset generation process, but not the actual data. Due to inaccuracies and ambiguities in these descriptions, it is often impossible to reproduce the dataset. For example, the referenced dataset from \citeauthor{Zhou.2013}~\cite{Zhou.2013} is not available, and that from \citeauthor{Schmicker.2019}~\cite{Schmicker.2019} is no longer publicly accessible. The dataset used by \citeauthor{Sahs.2012}~\cite{Sahs.2012} is poorly documented, while \citeauthor{Shabtai.2010}~\cite{Shabtai.2010} offer statistical data without additional information about the underlying apps. \citeauthor{LinaresVasquez.2016}~\cite{LinaresVasquez.2016} provide a list of APK names and corresponding labels, but do not include necessary features like the application versions or the complete dataset with all extracted features.} These issues are also raised by \citeauthor{Hutson.2018}~\cite{Hutson.2018}, who shows that AI is facing a reproduction crisis, with only about 6\% of authors publishing their code and less than a third disclosing their database.\footnote{Another factor for this could be the constant change and development of the extraction tools on the one hand and the apps on the other, which leads to inconsistent replications.} The absence of published code has negative effects on all areas of computer science, while the lack of available datasets is a particularly significant burden on research into modern AI systems, which require massive amounts of training data. The unavailability of code and data, and the resulting issues with reproducibility greatly impede research progress. 

To address these challenges, we have defined a series of requirements that must be met by future datasets. Some of these requirements extend beyond the domains discussed in this article and can be applied to other fields as well.

\begin{itemize} 

\item{\textbf{Including apps for different purposes:}} To identify malicious and cloned applications, or to detect minimal differences between the original and the pirated application, datasets are needed that contain the original applications, malicious applications, and pirated copies with malicious code. Developing systems that provide options for action requires datasets with features that enable the identification of similarity relationships between applications.

\item{\textbf{Providing multipurpose features and labels:}} To create a database for validation, replication, and further development of existing systems, the dataset must have a wide variety of features and labels that are useful for different purposes. Most existing datasets either only provide the app and the label or app features derived for a specific use case. These need to be expanded.

\item{\textbf{Research data management:}} Long-term, high-quality, and cross-institutional data management is required to ensure availability and timeliness. To be able to react to current developments and validate previously published works with the same dataset version, it is necessary to implement mechanisms that enable the data to be updated quickly and constantly. To ensure verifiability and replicability, it is essential to guarantee that all dataset versions are permanently available. 

\item{\textbf{Quality of the data sources:}} Future datasets must be based on knowledge from several sources like different app stores and malware sources to improve their reliability and completeness.

\item{\textbf{Documentation:}}
A continuous and conscientious documentation of all work processes including the data collection, data preprocessing and data usage is essential.

\end{itemize}

Generating such datasets requires a non-trivial feature extraction procedure. The following is an example of how an effective tool chain could be structured based on the tools and features identified in this article. At the beginning, APKs must be collected. All app sources mentioned in \cref{tab:Tab_2_Android_APK_Sources} should be considered. In addition to the APK, features such as app name, size, developer, rating, version, description, and app categories\footnote{In addition to its usage as a feature in malware detection, the app category can also serve as a label for similarity detection.} can be derived from the app stores. Existing databases such as the Android Malware Genome Project should also be used (see \cref{tab:Tab_2_Android_APK_Sources}). Web scrapers and crawlers can be used for this purpose.\footnote{See, for example, Selenium \cite{Selenium.20240423}, which can be used to automatically extract data from websites.} The data should then be prepared and harmonized. For example, APKs downloaded from app stores should be analyzed with a virus scanner so that this data can be tagged with appropriate malware labels. On the other hand, the data from the existing datasets should be enriched with missing information, such as metadata from the app stores. A comprehensive dataset can then be generated from features such as API calls, call graphs, strings, sensors, intents, permissions, functions, activities, services, and recipients. It is advisable to use different tools and combine their results to get a better end result. The generated dataset should be usable for training malware and similarity detection systems. All analysis steps must be meticulously documented and published along with all versions of the dataset.

\section{LIMITATIONS}
\label{sec:limitations}

This study looks at the features required for Android malware, clone, and similarity detection and provides a comprehensive overview of their acquisition and usability. The research and development of such systems requires a sufficient number of high-quality labels in addition to the above-mentioned features, at least for the validation of the results. 

In the case of Android malware detection\footnote{This also includes detecting piggybacked malware in clones. However, this process only requires labels for validation purposes.} and simple similarity detection, labels exist and have been used by several authors. In case of malware detection, results from virus scanners and labels from existing Android malware datasets such as Drebin or the Android Malware Genome Project (see \cref{tab:Tab_2_Android_APK_Sources}) can be used. For a simple similarity detection, app store categories can be used. The inclusion of both labels was proposed in the description of the schematic process for generating a comprehensive dataset (see section \ref{sec:creating_universal_interdisciplinary_dataset}).

However, high-quality labels that describe the apps' functionality in detail are required for recognizing functionally identical Android apps. However, no comprehensive dataset containing such labels is known. Some researchers solve this by using handwritten labels. However, this is expensive and time-consuming. Furthermore, none of the reviewed sources take into account that modern Android apps can be multifunctional and therefore need to be assigned to multiple labels. These problems need to be addressed in future research. A starting point could be the development of a partly automated labeling process or a crowdsourcing approach.

\section{CONCLUSION}
\label{sec:conclusion}

This paper analyzes the literature on Android malware detection, clone detection, and functional similarity detection between 2002 and 2022, focusing on the underlying features. We extracted and analyzed data sources, data collection tools, and feature mining processes from scientific publications. In doing so, we lay the foundation for the creation of a publicly available, universal and interdisciplinary Android feature dataset. 

The evaluation of the examined literature showed that a wide range of sources, methods and tools are available. A critical step in collecting features from Android apps is to collect the app's APK, which can be obtained from app stores and public malware sources. Another important step is feature extraction. To extract features from both the marketplace and the app's APK, several tools and toolchains were analyzed, including those for decompiling and analyzing the APK, examining the underlying Java code, and generating statistics and graphics. The analyzed features showed that textual features such as app description, author information, and app ratings can be easily obtained by crawling Android marketplaces, but according to \citeauthor{Pandita.2002}~\cite{Pandita.2002}, they are more susceptible to manipulation. In contrast, technical features extracted directly from the APK are more tamper-proof and reliable for security-related analysis due to their direct association with the application. However, their extraction is much more difficult due to their complexity. The paper also shows that the most commonly used features of the APK are API calls, permissions, and code analysis. Since code is at the heart of any application, techniques such as extracting and analyzing diagrams, n-grams, and statistics, or using hashing and filtering techniques to reduce the search space have been extensively explored in this article. Metadata, such as the internal structure of the application, and textual features, such as strings, are used less frequently.

We explored how to make the extracted features useful for malware, clone, and similarity detection purposes, aiming to acquire profound insights into feature utilization and uncover potential synergies. It is described that synergies can be exploited by extracting overlapping features in these domains. In this context, the work examines the symbiotic relationship between these areas and how they can benefit from each other. Malware detection systems can benefit from clone detection systems because cloned applications often contain malicious code. Clone detection systems, in turn, can benefit from the technical concepts of (functional) similarity detection systems, which in turn provide users with solutions to risks identified by malware and clone detection systems.

An interdisciplinary Android dataset is vital for research as it enables validation, replication, and the development of new methods and models. Its versatility allows it to address various problems across disciplines, fostering general and robust solutions and opening avenues for new research fields. Additionally, using an existing dataset reduces the time and effort required for data collection, enabling researchers to work more efficiently. This paper also highlights the importance of the human factor in mitigating security risks. For example, inexperienced users may not act appropriately despite warnings from malware or clone detection systems, posing a potential risk to the entire Android ecosystem. It is therefore essential to have combined systems that inform users about dangerous applications and provide them with options for action to protect themselves.

In summary, our analysis shows that there is a significant lack of publicly available, universal, and interdisciplinary Android feature datasets that can be used in the analyzed research areas. This lack impedes progress and collaboration in the field of Android security research. To address this issue, we outline the requirements and the process of creating such a publicly available, universal, and interdisciplinary Android feature dataset. Authors need to 1) Include apps for different purposes, 2) Provide multipurpose features and labels, 3) Implement and maintain research data management, 4) Ensure quality of the data sources, and 5) Document adequately.

In addition to the researchers, publishers can also help to ensure availability and reproducibility of research outcomes. For example, journals could require a reference to a publicly accessible dataset or the publication of the dataset used, along with the publication of documentation and source code on the journal's website alongside the papers. This would enable the broader research community to reproduce and assess the findings of the published work. A freely available dataset, such as the one proposed in this article, can be helpful in this regard.

The review not only summarizes the current state of research but also underscores the importance of collaborative efforts to overcome existing challenges. The promotion of open data-sharing practices and the encouragement of reproducible research can facilitate the advancement of Android security research and empower users to make informed decisions about app security.

\section{ACKNOWLEDGMENTS}
\label{sec:acknowledgments}

%We would like to thank our colleagues at the DAI-Laboratories of the TU Berlin for their helpful feedback and support. In particular, We would like to thank Karsten Bsufka for his valuable contributions to this research.
%This work is funded by the German Federal Ministry of Education and Research (BMBF) under grant number 16SV8518.
\textit{<blinded>}

\printbibliography
\appendix
\newpage
\begin{landscape}

\section{Appendices}
\label{sec:Appendices}

\begin{longtable}{|p{2.5cm}|p{2cm}|p{2cm}|p{3.5cm}|p{5cm}|p{7.5cm}|}
    \caption{Extended feature extraction tools}\\
    \label{tab:Tab_3_Feature_extraction_tools_extended}\\
    \hline    
    \textbf{Category} & \textbf{Name} & \textbf{Paper} & \textbf{Input} & \textbf{Output} & \textbf{Usabillity}\\
    \hline
    \endfirsthead
    \hline
    \multicolumn{6}{|c|}{{\bfseries Table \thetable\ continued from previous page}} \\
    \hline
    \textbf{Category} & \textbf{Name} & \textbf{Paper} & \textbf{Input} & \textbf{Output} & \textbf{Usabillity}\\
    \hline
    \endhead
    \hline
    \multicolumn{6}{|c|}{{Continued on next page}} \\
    \hline
    \endfoot
    \hline
    \endlastfoot

            Programming language & python \cite{Python.org.20240404} & \cite{Chen.2014} & Source code files & -  & Provides a scripting language for automation and tool development.\\
            
            \hline
            
            APK decompiler & apktool \cite{Tumbleson.2020} & \cite{Anupama.2022, Batyuk.2011b, Gorla.2014, Hosseini.2021, LinaresVasquez.2016, LinaresVasquez.2014, Shao.2014, Sun.2014, Tian.2015b, Wang.2015, Xu.2018, Zhang.2018} & Android APK file &  decompiled files (dex, jar, java, smalie, ...), resources (xml, jpg, ...), manifest file  & Assemble/Disassemble APK files, aiding in analyzing code for malicious patterns and identifying clones.\\
            
            APK decompiler & Baksmali \cite{Baksmali.20240404} & \cite{Blasing.2010, Chen.2014, McLaughlin.2017, Shao.2014, Sun.2014, Zhou.2013, Zhou.2012} & Dalvik bytecode (dex) & Smali code (human-readable assembly-like code) & Generates readable smali code, which can be analyzed for malicious behaviors.\\
            
            APK decompiler & Dedexer \cite{Paller.2009} & \cite{Felt.2011, Sanz.2012} & Dalvik bytecode (dex) & Decompiled source code, smali code & Helps in decompiling DEX files for understanding app logic and identifying malicious activities.\\
            
            APK decompiler & dexdump \cite{archlinux.06.09.2022} & \cite{Potharaju.2012} & Dalvik bytecode (dex) & Decompiled source code, various information about DEX file & Provides detailed information about DEX files, aiding in understanding app structure and identifying anomalies.\\
            
            APK decompiler & dex2jar \cite{Pan.2016} & \cite{Crussell.2012, Crussell.2013, Crussell.2015, LinaresVasquez.2016, Shao.2014, Tian.2015b, Wang.2015, Yang.2014} & Dalvik bytecode (dex) & Java bytecode (JAR file) & Converts DEX files to JAR format for further analysis, enabling detection of malicious code patterns.\\
            
            \hline
            
            APK analysis tool & Androguard \cite{Desnos.2018} & \cite{Crussell.2012, Sahs.2012,Wang.2015,Yang.2014,Zhang.2015} & Android APK file & Various metadata including permissions, API calls, intents, activities, services, broadcast receivers, and more. & Analyzes APKs for suspicious behaviors, permissions, and API calls, aiding in malware detection.\\

            APK analysis tool & A3E \cite{Azim.2013} & \cite{Shao.2014} & Decompiled source code, bytecode, or APK file & static and dynamic analyze, resource extraction, Control flow graph, call graph & Analyze apps and extract resources and graphs, aiding in visualizing app structures and behaviors.\\
            
            APK analysis tool & Keytool \cite{Oracle.2022} & \cite{Wang.2015, Zhou.2012} & Android APK file & Certificate information, key information, signature details. & Extracts certificate information for verifying app authenticity, helping to identify potentially malicious apps.\\
            
            APK analysis tool & Soot \cite{ValleeRai.1999} & \cite{Chen.2013} & Android APK file, .jar, .class & Intermediate representation (Jimple, Shimple, Grimp), control flow graph, call graph, points-to analysis results. & Provides control flow graphs and call graphs for app analysis, aiding in identifying code similarities and anomalies.\\

            APK analysis tool & aapt \cite{AndroidDevelopers.25.08.2020} & \cite{Anupama.2022, Schmicker.2019, Shao.2014} & Android APK file & Various information including package name, version code, version name, permissions. & Injects and extracts information and data, aiding in understanding app properties and permissions.\\
            
            \hline
            
            Java analysis tool & BCEL Java Library \cite{ApacheSoftwareFoundation..2020} & \cite{Tian.2015b} & Java bytecode (jar and class files) & Parsed bytecode instructions, class structure information. & Analyzes Java bytecode, aiding in identifying code vulnerabilities and detecting malicious code patterns.\\
            
            Java analysis tool & ckjm \cite{Spinellis.2005} & \cite{Tian.2015b} & Java bytecode (class files) & Different complexity metrics. & Calculates cyclomatic complexity, weighted methods per class, and lack of cohesion in methods aiding in identifying complex or potentially obfuscated code.\\
            
            Java analysis tool & JAD Compiler \cite{Kouznetsov.2001} & \cite{LinaresVasquez.2014} & Java bytecode (class files) &  Decompiled source code (Java files). & Decompiles Java bytecode into readable source code, aiding in identifying malicious patterns.\\
            
            Java analysis tool & JADX Compiler \cite{Skylot.2020} & \cite{Hosseini.2021} & Android APK file, .jar, .class, .dex &  Decompiled source code (Java files). & Decompiles Java bytecode into readable source code, aiding in analyzing code for malicious patterns.\\
            
            Java analysis tool & JClassInfo \cite{Anarxia.2004} & \cite{LinaresVasquez.2016} & Java bytecode (jar and class files) & Class hierarchy information, method signatures. & Provides information about class hierarchy and method signatures, aiding in code analysis.\\
            
            Java analysis tool & JD Core \cite{GitHub.08.09.2022} & \cite{Wang.2015} & Java bytecode (jar and class files) & Decompiled source code (Java files). & Decompiles Java bytecode into readable source code, aiding in analyzing code for malicious patterns.\\
            
            Java analysis tool & WALA \cite{IBM.2022} & \cite{Crussell.2012,Crussell.2013,Crussell.2015, LinaresVasquez.2014} & Java bytecode (jar, class, and aar files) & Intermediate representation (IR), control flow graph, call graph. & Provides an intermediate representation for Java bytecode, aiding in analyzing complex code structures.\\
            
            Java analysis tool & RAPID library \cite{Zhang.2016b} & \cite{Schmicker.2019} & .dex, .apk & Structural analyse of input data. & Analyze results and summaries, aiding in code analysis.\\

            \hline
            
            Generic analysis tool & IDA \cite{Hex-Rays.2022} & \cite{Hosseini.2021} & Different executables (exe, dll, apk, ...) & disassembly, code analysis, data structure views, dependency graphs & Provides analysis results, interactive views and dependency graphs, aiding in deep analysis of app structures.\\
            
            \hline
            
            Hash Tool & LSH code \cite{Andoni.2006} & \cite{Crussell.2013, Crussell.2015} & Data to be hashed & Locality Sensitive Hash (LSH) values & Computes Locality Sensitive Hash (LSH) values, aiding in identifying similarities.\\
            
            Hash Tool & MinHash \cite{Broder.1997, Broder.2000} & \cite{Crussell.2013, Crussell.2015} & Data to be hashed & MinHash signature & Generates MinHash signatures, aiding in identifying similarities.\\

            \hline
                    
            Helper Tool & 7zip \cite{Riehm.16.07.2022} & \cite{LinaresVasquez.2016, LinaresVasquez.2014} & De-/Compressed files & De-/Compressed files & De-/Compress files, aiding in app packaging and analysis.\\
            
            \hline
            
\end{longtable}

\end{landscape}

% \begin{acks}
% The authors are grateful to the editors of DTRAP and welcome feedback from readers and collaboration from the community.
% \end{acks}
%\appendix
%\input{sections/99_Appendix/99_Appendix}

\end{document}